\documentclass[]{article}
\usepackage[normalem]{ulem}
\usepackage[margin=1in]{geometry}
\usepackage{comment}

\usepackage[utf8]{inputenc}
\usepackage{newtxtext}
\usepackage{amsmath}
\usepackage{amssymb}
\usepackage{newtxmath}

\usepackage[pagebackref = false,]{hyperref}
\usepackage{color}
     \usepackage{xcolor}
\usepackage{epsfig}
\usepackage{feynmp-auto}
\usepackage{multirow}
\usepackage{amsfonts}
\usepackage{amssymb}
\usepackage{graphicx}
\usepackage{tabularx}
\usepackage{array}
\usepackage{outlines}
\hypersetup{
    colorlinks = true,
    urlcolor   = blue,
    citecolor  = black,
}
\usepackage{bm}

\usepackage{wrapfig}
\usepackage[font=small]{caption,subcaption}

\def\bk{\mathbf{k}}

     \usepackage{color}
     \usepackage{xcolor}
\definecolor{g-blue}{rgb}{0.83,0.95,1}
\definecolor{g-yellow}{rgb}{1,1,0.7}
\definecolor{g-green}{rgb}{0.9,1,0.9}
\definecolor{green}{rgb}{0,0.6,0}
\definecolor{cyan}{rgb}{0,0.7,0.7} 
\definecolor{grey}{rgb}{0.4 ,0.4 ,0.4 }
\definecolor{brown}{rgb}{0.6 ,0  ,0.8 }
\def\brown#1{\textcolor{brown}{#1}}
\def\g-blue#1{\textcolor{g-blue}{#1}}  

\def\black#1{\textcolor{black}{#1}}
\def\blue#1{\textcolor{blue}{#1}}

\usepackage[normalem]{ulem}
\usepackage{chngcntr}
\usepackage[mathscr]{euscript}

\usepackage{leftindex}
\usepackage{etoolbox}

\usepackage{titlesec}
\usepackage[style=apa, backend=biber]{biblatex}
\hypersetup{
colorlinks=true, 
linkcolor=black, 
citecolor=black, 
filecolor=black, 
urlcolor=black 
}

\title{Spectral Transfers of Sign-Definite 
Invariants in Wave Turbulence}

\begin{document}

\numberwithin{equation}{section}

\newcommand{\cale}[0]{\mathcal{E}}
\newcommand{\cali}[0]{\mathcal{I}}
\newcommand{\ucale}[0]{\underline{\cale}}
\newcommand{\fp}[1]{\leftindex[I]^{#1}{P}}
\newcommand{\ff}[1]{\leftindex[I]^{#1}{F}}
\newcommand{\bx}{\textbf{x}}
\newcommand{\calj}{\mathcal{J}}
\newcommand{\calr}{\mathcal{R}}
\newcommand{\sgn}{\text{Sgn}}
\newcommand{\qdensity}{\mathcal{Q}}
\newcommand{\Qinteract}{{{\dot Q}}}

\def\stackbelow#1#2{\underset{\displaystyle\overset{\displaystyle\shortparallel}{#2}}{#1}}

\title{Spectral Transfers of Sign-Definite
Invariants in Wave Turbulence}

\author{Nicholas R. Salvatore$^{1}$, Giovanni Dematteis$^{2,3}$, and Yuri V. Lvov$^{4}$\\
$^{1}$Naval Surface Warfare Center, Philadelphia Division\\
$^{2}$Physics Department, Universit\`a degli Studi di Torino\\
$^{3}$Physical Oceanography Department, Woods Hole Oceanographic Institution \\
$^4$Department of Mathematical Sciences, Rensselaer Polytechnic Institute, Troy, NY }
\date{\today}
\maketitle
\

\abstract{
We consider the spectral transfers of sign-definite invariants in general wave turbulence systems equipped with a wave kinetic equation. 
We develop a formalism to investigate and to characterize these transfers, based on the ability to track the exchanges within individual interactions by the detailed-conservation properties of the resonant interactions.  
We consider the three-wave, four-wave, five-wave and six-wave kinetic equations, introducing measures of locality of general wave turbulence system and of
median step size in wave-wave scattering. We then use the proposed method to study the
Majda-McLaughlin-Tabak one-dimensional model and examine the characteristics of energy and action transport. We show that  energy in the direct energy cascade is dominated by local interactions, but shows a nonlocal inverse ``return flow'' that counters the direction of the cascade. The inverse action cascade is instead ruled by more nonlocal interactions, but shows a highly local direct ``return flow''. We also show that for both cascades a scale separation of over an order of magnitude is necessary to ensure a negligible interaction rate with forcing or dissipation regions, hinting at why in numerical simulations of the model an extremely large box size is necessary in order to realize the theoretical Kolmogorov-Zakharov spectra.}
\section{Introduction}

Wave turbulence is a theoretical framework used to describe the systems of weakly interacting particles or waves with a large number of degrees of freedom. Examples of such systems include surface gravity waves, surface capillary waves, internal waves in the ocean, and atmospheric waves to make just a few examples. 
We refer the reader to
\parencite{zakharov2012kolmogorov,NazBook,galtier2022physics}. 

The focal point of the wave turbulence theory is the wave kinetic equation describing the time evolution of the spectral energy density of a system. The wave kinetic 
equation assumes a different form depending on the possible resonances between interacting waves. For example, in three-wave systems the spectral transfers happen between triads of wave numbers, in four-wave systems between quartets, and so on. All the possible interactions between $n$-plets of wave numbers in resonance with a given wave number are accounted for in the  collision integral, where they are all summed together giving as a result the time rate of change of the wave action of the given wavenumber.
Examples of three-wave
systems include surface capillary waves \parencite{Zakharov1967capillary,Pushkarev:96} and internal waves in the ocean \parencite{muller1975dynamics,lvov2001internal,LT2}; the typical exaple of a four
wave system is given by surface gravity waves \parencite{hasselmann62,zakharov1968stability,komen1996dynamics,resio1991numerical}, the five wave systems are one-dimensional gravity waves
\parencite{Krasitskyfivewave,DYACHENKOLVOVZAKHAROV1995FiveWave,L}. An example of a six-wave system is the celebrated $\alpha$-Fermi-Pasta-Ulam-Tsingou system \parencite{Onorato2015PNAS,ONORATO20231} with a moderate number of degrees of freedom. 

Remarkably, the Kolmogorov-Zakharov spectrum is a
    steady-state solution of the wave kinetic equation. For these solutions to be physically realizable, the collision integral has to be convergent. Usually, a solution to the wave kinetic equation is said to be  ``local" if the 
collision integral of the kinetic equation 
converges, and ``nonlocal'' if the collision integral diverges. 
This binary classification of locality
was improved upon by \cite{dematteis2023structure}, who introduced a locality coefficient, describing the degree to which the system is local or nonlocal.  In \cite{dematteis2023structure}  the three-wave systems were considered,
like oceanic internal waves or surface capillary
 waves. 

The purpose of this manuscript is to develop
a general theory that covers not only three-wave resonances, but also four-, five- and six-wave resonances. We introduce locality metrics that characterize how local the 
interactions between different wave numbers are.
We also introduce a 
characteristic ``step size", i.e. the typical median length (similar to a mean free path) of spectral interactions in the wave number space. 

Specifically, we consider the spectral transfer structure of wave kinetic equations. We  consider four cases, where the number of interacting waves can be an integer $n$ between three and six. The common feature of these wave kinetic equations is that they have a delta function of differences of frequencies which enforces the conservation of the quadratic energy in each individual interaction of $n$ wave modes. We assume that the quadratic energy is very close to the full energy of the pre-averaged system, and as such this conservation serves as a close approximation to the conservation of total energy in individual interactions.

We apply our method to the celebrated 
MacLaughlin-Majda-Tabak model as a prototype wave turbulence system to gain information on the transfers of conserved quantities across the wave number space.

The paper is organized in the following way. In sections (\ref{Theory_fourwave}-\ref{Theory_others}) we develop a universal theory 
of locality of spectral transports for three-, four-, five, and six-wave systems. We establish the 
necessary formalism to characterize such interactions. In Section (\ref{Numerics}) we describe the results of our formalism applied to the Majda-McLaughlin-Tabak  model. 
We conclude in section (\ref{Conclusions}). 

\section{The Four-Wave Transfer Rate Formula
}\label{Theory_fourwave}

In this section we build a theory that tracks these individual scattering events to calculate the total 
amount of a quadratic invariant (such as energy) that is transferred between two disjointed sets in wave number space. We first consider the four wave kinetic equation with two-to-two scattering interactions, which serves as a generic form of four wave turbulent systems.

In the next section, we further develop this theory for various wave kinetic equations
for three, five, and six interacting waves. 
Our formalism is general regardless of the specific invariant tracked, be it energy, action (e.g. in four-wave systems), enstrophy (e.g. see Rossby waves), or other quantities depending on the system at hand.

\subsection{The Four Wave Kinetic Equation as an Interaction System}\label{SectionFourWave_derivation}
\subsubsection{Properties of the four-wave kinetic equation}
We start by considering the four-wave kinetic
equation for the case when two incoming waves are 
scattered into two outgoing waves (two-to-two wave kinetic equation) 
\parencite{ZLF, NazBook,  galtier2022physics}:

\begin{equation}\label{KineticEqnFourWave}
\begin{aligned}
\dot{n}_\textbf{k}&= \int d\textbf{k}_{123}\cali^{12}_{3\bk}\,,\qquad \text{where}\\
\cali^{12}_{3\bk}=4\pi|V^{12}_{3\textbf{k}}|^2I^{12}_{3\textbf{k}}\Delta^{12}_{3\textbf{k}}\,,&\qquad I^{12}_{3\bk} =
n_1n_2n_3n_\bk\left(\frac{1}{n_\bk}+\frac{1}{n_3}-\frac{1}{n_1}-\frac{1}{n_2}\right)\,,\\ 
\Delta^{12}_{3\bk}=\delta(\bk_1&+\bk_2-\bk_3-\bk)\delta(\omega_1+\omega_2-\omega_3-\omega_\bk)\,.
\end{aligned}
\end{equation}
The right-hand side of the wave kinetic equation is 
traditionally called the collision integral.
The wave action is denoted as $n_\bk$, and it is a spectral energy density divided by the linear dispersion $\omega_\bk$ . Furthermore, the  $V^{12}_{3\textbf{k}}$  represents the interaction matrix element, i.e. the strength of the nonlinear interactions between the wave numbers ${\bf k}$, ${\bf{k_1}}$, ${\bf{k_2}}$ and ${\bf{k_3}}$. 
The beauty of the wave turbulence theory is that the values of the matrix element change from problem to problem, and the kinetic equation form stays the same.

{We assume that the matrix elements satisfy the following properties (e.g., for Hamiltonian systems this is necessary for the Hamiltonian to be real valued):
\begin{equation}
(V^{3\bk}_{12})^* =V^{12}_{3\textbf{k}} = V^{21}_{3\textbf{k}}=V^{12}_{\textbf{k3}}\,.
\end{equation}

Here, the ``$*$'' symbol denotes complex conjugation. Then, we have:
\begin{equation}
\cali^{12}_{3\textbf{k}} = \cali^{12}_{\textbf{k3}} = - \cali_{12}^{3\textbf{k}} = - \cali_{21}^{3\textbf{k}}\,.\label{SymmetriesOfI}
\end{equation}
It is interesting to note that the structure of the kinetic equation
    kernel has the following property: if there is a four-wave
    interaction of wave numbers ${\bf{k_1}}$ and ${\bf{k_2}}$ colliding to
    scatter into wave numbers ${\bf{k_3}}$ and ${\bf{k_4}}$, then the amount of wave action lost
    by ${\bf{k_1}}$ and  ${\bf{k_2}}$ are equal to each other.
    Furthermore, amount of wave action gained by ${\bf{k_3}}$ and ${\bf{k_4}}$
    are also equal to each other.
}
Simultaneous fulfillment of the two delta functions in $\Delta^{12}_{3\bk}$ enforces the \textbf{resonance conditions} for the resonant wave quartets, and we refer to the family of quartets which conform to these resonance conditions as the \textbf{resonant manifold}

\begin{equation}\label{ResonantManifoldFourWave}
\bk_1+\bk_2=\bk_3+\bk\,,\quad \omega_1+\omega_2=\omega_3+\omega_\bk\,.
\end{equation}

Due to the structures of resonant manifold  and the symmetries of the collision integrand, the four wave kinetic equation \eqref{KineticEqnFourWave} has two positive-definite integrals of motion. These are the total number of particles {(or total wave action)}  $N$ and the quadratic energy $E$ defined respectively as
\begin{equation}\label{basicintegralconservation}
N=\int d\bk\:n_\bk;\quad E=\int dk\:\omega_\bk n_\bk
\end{equation}

It turns out that 
conservation \eqref{basicintegralconservation} is a reflection of a deeper conservation in the collision intgrand referred to as detailed conservation or detailed balance ~\parencite{onsager1949statistical,kraichnan1959structure,hasselmann1966feynman,rose1978fully,eyink1994energy}. This observation allows us to generalize  the notion
of conserved quantity. 
Let $Q_\bk = \rho_\bk n_\bk$ represent a quadratic density. By the resonance conditions \eqref{ResonantManifoldFourWave}, $\qdensity=\int d\bk Q_{\bk}$ is a constant of motion if 

\begin{equation}\rho_\bk+\rho_3=\rho_1+\rho_2,\:\:\rho_\bk\geq 0\label{resonantrho}.\end{equation}
A quantity $\rho_k$ satisfying Eq.~\eqref{resonantrho} is called a {\bf collision invariant}.
Note that if $\rho_\bk=1\text{ or }\omega_\bk$, then  $\qdensity$ is the total wave {action} $N$ or quadratic energy $E$, with $1$ and $\omega_\bk$ satisfying Eq.~\eqref{resonantrho}. 
Now, consider a resonant quartet $\bk_1,\:\bk_2,\:\bk_3,\:\bk$ that satisfies the resonance conditions \eqref{ResonantManifoldFourWave}. We then introduce a weighted collision integrand, which we call \textbf{{detailed} interaction rate} 

\begin{equation}\label{IndividualInteractionRate}
{\dot Q}^{123}_\bk = {\rho_\bk}\cali^{12}_{3\textbf{k}};\quad \dot Q_\bk =\int d\bk_{123}\Qinteract^{123}_{\bk}.
\end{equation}

We interpret ${\dot Q}^{123}_\bk = {\rho_\bk}\cali^{12}_{3\textbf{k}}$ as the spectral density interaction rate due to the interaction of $\bk$ via the considered resonant quartet with the other three modes $\bk_1,\bk_2, \bk_3$. Such notation allows us to "single out" a variable of the interaction. We  therefore consider $Q^{123}_{\bk}$ to be the contribution of the individual interaction of this quartet to the evolution of $Q_\bk$, and likewise we may consider $\Qinteract^{3\bk 2}_{1}$ to be the contribution of this individual quartet interaction to the evolution of $Q_1$. Using this notion, we can both present the detailed conservation relationship, and pave the way for its subsequent interpretation. The \textbf{detailed conservation}, or \textbf{detailed balance} relationship ~\parencite{onsager1949statistical,kraichnan1959structure,hasselmann1966feynman,rose1978fully,eyink1994energy, dematteis2023structure} is given by
\begin{equation}\label{DetailedBalance}
    {\dot Q}^{123}_\bk + {\dot Q}^{12\bk}_3 = - {\dot Q}^{3\bk1}_2  - {\dot Q}^{3\bk2}_1\,.
\end{equation}

We previously referred to the detailed conservation as a ``deeper conservation'' than the fact $\qdensity=\int d\bk Q_\bk$ is a constant of motion. We see that the general conservation of $Q$ follows directly from the detailed conservation, as it encapsulates both the (skew)symmetries of the collision integrand, and the condition \eqref{resonantrho} on $\rho_\bk$. Ultimately, rather than simply observing a total (integral) conservation of $Q$, we find every wave interaction contributing to the evolution of $Q_\bk$ must \textit{itself} be conservative, in the sense of \eqref{DetailedBalance}.

Thus, we interpret the equations \eqref{SymmetriesOfI}-\eqref{DetailedBalance} in analogy with rates of particle interaction: (i) the mode variable $\bk$ labels a ``wave state,'' described by spatial and temporal period, and the action density $n_\bk$ allows to ``count the number of waves'' in such a state (in a continuum sense, as a density). The rate at which these wave states interact with each other balance such that as pairs of ``incoming'' waves (the pair with negative interaction rate) ``collide,'' exchanging energy and action, they scatter into pairs of ``outgoing'' waves (the pair with positive interaction rate), preserving their total ``number'' (action) and their total energy. The wave kinetic equation ~\eqref{KineticEqnFourWave} sums all the interaction rates between the different wave states which resonate with $\bk$. This interpretation has given rise to extremely fruitful formal analogies with the Boltzmann equation for a gas of particles and with quantum field theory \parencite{peierls,hasselmann62,spohn2006phonon,ONORATO20231} 
with huge advantages, for instance, in the application of diagrammatic techniques that are still being used today to investigate the well-posedness of wave kinetic equations. \parencite{choi2004probability,eyink2012kinetic,chibbaro20184,deng2021full,chibbaro2017wave}

\begin{figure}
    \centering
    \includegraphics[width=0.5\linewidth]{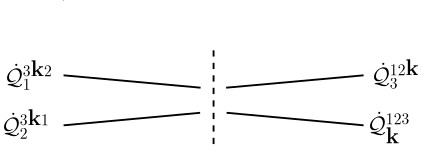}
    \caption{Scattering diagram interpreting detailed balance schematically. 
    }
\label{fig:ScatteringDiagram}
\end{figure}

Hence, for a fixed resonant quartet we slot the interaction rates into a scattering diagram, with the pairs $(12)$ and $(3\bk)$ representing sides of the scattering relationship (fig \ref{fig:ScatteringDiagram}). We treat the interaction rates as interacting wave states, as the interaction rate is uniquely indexed by the wave state, and the interaction rates are given by the ensemble average of interacting wave states from the model equation. We do not direct the diagram as the sign of the interaction rate $\Qinteract^{123}_{\bk}$ is itself undetermined at this stage: it is the relative sign of the interaction given by the skew symmetry of the collision integrand \eqref{SymmetriesOfI} and the detailed balance \eqref{DetailedBalance} which is relevant in our construction.

We note that the reasoning below can be further generalized. For example, the wavenumber delta function in $\Delta_{3\bk}^{12}$ ensures the existence of additional invariant for $\rho_\bk=k_j$, where $j$ indicates the wavenumber components. These extra invariants are the components of the  quasi-momentum, making the analogy with particles even deeper. However,   since these extra invariants are not sign-definite, they are not
considered here. 

\subsubsection{Derivation of the Four Wave Transfer Rate Formula}

We now construct the four wave transfer rate formula for conserved density $Q_\bk$ from subset $A$ to subset $B$ of the wave number domain.  
Consider the evolution of the invariant density $Q_\bk$ over a subset $A\subset\mathbb R^n$ of the wave number domain with respect to the four wave kinetic equation ~\eqref{KineticEqnFourWave},

\begin{equation}\label{EnChangeA}
\partial_t\int_A d\bk\: Q_\bk=\int_A d\bk\: \dot{Q}^{123}_\textbf{k}= \int_A d\bk\: \rho_\bk\int d\textbf{k}_{123}\:\cali^{12}_{3\bk}\,.
\end{equation}

We will first extend the (skew)symmetries of the collision integrand \eqref{SymmetriesOfI} to an equivalent collection of relationships for the spectral interaction rates \eqref{IndividualInteractionRate}. We see that these symmetries lead to a collection of relationships which describe the relative contributions of each wave state density to the detailed balance relationship \eqref{DetailedBalance} of the form

\begin{align}
\frac{1}{\rho_1}{\dot Q}^{3\bk2}_{1}=\frac{1}{\rho_2}{\dot Q}^{3\bk1}_{2}=-\frac{1}{\rho_3}{\dot Q}^{12\bk}_{3}=-\frac{1}{\rho}{\dot Q}^{123}_{\bk}\,.\label{EquivQ}
\end{align}

We consider four categories of interaction, fixing pairs $(12)$ and $(3\bk)$ and only permuting the relative relationship between $\bk$ and the other three fixed wave numbers. We could break the interaction rate and detailed conservation over all permutations of the scattering diagram; however, this would not provide additional information in this case.
In the five wave interaction case considered below the situation will be different. 

We interpret the detailed conservation form and associated properties~\eqref{DetailedBalance}-\eqref{EquivQ} as follows: We schematically place the densities with shared signs (by \eqref{SymmetriesOfI} and \eqref{resonantrho}) on the same side of the annihilation-scattering interaction diagram (Fig. \ref{fig:ScatteringDiagram}), with ${\dot Q}^{3\bk2}_{1}$, ${\dot Q}^{3\bk1}_{2}$ being the ``input", and ${\dot Q}^{12\bk}_{3}$, ${\dot Q}^{123}_{\bk}$ being the ``output,'' with the direction (determining annihilation vs scattering) of the interaction left ambiguous. We have fixed $\bk \in A$, and are only interested in counting scattering operations which relate the interaction rate with respect to $\bk$ to the interaction rates of the ``input'' wave states $\bk_1$ and $\bk_2$. We find through the ratio relationships of \eqref{EquivQ} that the output interaction rates indexed by $\bk$ and $\bk_3$ can be treated independently due to the symmetries of the collision integrand. Hence, we express the detailed conservation as a relation between the interaction rate of the wave $\bk$ under consideration, and the interaction rates with respect to the scattered/annihilated wave states $\bk_1$ and $\bk_2$ on the other end of the interaction. 

\begin{equation}
- {\dot Q}^{3\bk1}_2  - {\dot Q}^{3\bk2}_1 = \frac{\rho_\bk+\rho_3}{\rho_\bk}{\dot Q}^{123}_\bk
\end{equation}

Leveraging the 
property \eqref{resonantrho} of $\rho_\bk$ and substituting into the change in $Q_\bk$ over set $A$ \eqref{EnChangeA}, we have:

\begin{equation}
\int_A d\bk \dot{Q}_\textbf{k}=- \int_A d\bk \int d\textbf{k}_{123}\left[\frac{\rho_\bk}{\rho_1+\rho_2}{\dot Q}^{3\bk2}_1+\frac{\rho_\bk}{\rho_1+\rho_2}{\dot Q}^{3\bk1}_2\right]
\end{equation}

This equation expresses the change in the spectral density inside a set $A$ via contributions from colliding waves that scatter into/out of wave $\bk$. To further generalize, we restrict interaction wave numbers based upon their set of origin/destination. Specifically, We only count individual contributions
to spectral density of $\bk\in A$ that originate from/decay into waves in a set $B$. Thus, the formula for the total transfer from set $A$ to set $B$ becomes

\begin{equation}
\fp{\rho}_{A\to B}^4 = \int_A d\bk \int d\textbf{k}_{123}\left[\frac{\chi_B(\bk_1)\rho_\bk}{\rho_1+\rho_2}{\dot Q}^{3\bk2}_1+\frac{\chi_B(\bk_2)\rho_\bk}{\rho_1+\rho_2}{\dot Q}^{3\bk1}_2\right]\,,
\end{equation}
where
\begin{equation}
    \chi_B(\bk) = \left\{\begin{array}{cc}
       1\,,  &  \text{if } \bk \in B\,,\\
       0\,,  &  \text{if } \bk \notin B \,.
    \end{array}\right.
\end{equation}
Using the notation we introduce above, this reduces to 

\begin{equation}\label{fourwaveflux}
\fp{\rho}_{A\to B}^4=- 2\int_A d\bk \int_B d \bk_1 \int d\bk_{23}\frac{\rho_\bk \rho_1}{\rho_1+\rho_2}\mathcal I^{12}_{3\bk}\,.
\end{equation}

The equation \eqref{fourwaveflux} represents one of the main results of this paper. It allows us to calculate explicitly the rate at which conserved density $Q_\bk=\rho_\bk n_\bk$ flows between two disjoint sets A and B. We show in the following how such a characterization allows for the calculation and visualization of generalized fluxes in the four-wave systems. 
{For brevity, we refer to this equation as the transfer formula, and call $^\rho P_{A\to B}^n$ the transfer rate between set $A$ and $B$, relative to the collision invariant $\rho$, and due to a collision integral with $n$-wave interactions.}

\subsubsection{Properties}

{The transfer formula \eqref{fourwaveflux} enjoys the following properties, which are quite intuitive for fluxes of conserved quantities in physical systems (we drop the subscripts $\rho$ and $n$ for ease of notation, as these properties are valid for any collision invariant $\rho$ and for any number $n$ of interacting waves):} 
\begin{align}
&P_{A\cup B \to C}=P_{A\to C}+P_{B\to C}-P_{A\cap B\to C}\label{prop1leftsplit}\\
&P_{A\to B\cup C}=P_{A\to B}+P_{A\to C}-P_{A\to B\cap C}\label{prop2rightsplit}\\
&P_{A\to B}=-P_{B\to A}\label{prop3flip}\\
&P_{A\to A}=0\label{prop4selfinteractions} 
\end{align}

The last property \eqref{prop4selfinteractions} follows from the property \eqref{resonantrho} of collision invariant $\rho_\bk$ and will be detailed in the following section.\\

\subsection{Conservation Form of the Wave Kinetic Equation and Scale Invariant Solutions}\label{sectionfluxmathchup}

{In this section, we illustrate the connection between the transfer formula \eqref{fourwaveflux} and the spectral flux that is usually computed in the wave turbulence literature.}
Let us assume spatial isotropy, and denote by $\Omega$ the solid angle, with the integral implicitly performed over the {$d$-dimensional} unit sphere. Then the wave kinetic equation can be expressed by mapping to $d$-spherical coordinates:
\begin{equation}\label{isotropicfirsttime}
\dot n_k =\int_0^{\infty} dk_{123} d\Omega_{123} \mathcal I^{12}_{3k}
\end{equation}
When $n_k$ is a function of a single parameter $k=|\bk|$, we may, by the Leibniz integral rule, write the evolution of a conserved density $\rho_kn_k$ by \eqref{isotropicfirsttime} in a continuity equation form: \parencite{ZLF, NazBook} (Note the transition to the variable $K$ is a disambiguation to delineate it from the variable of integration $k$ in \eqref{classicalflux})

\begin{equation}\label{ConservationForm}
\rho_K\dot n_K = -\partial_K\;{^{\rho}P_K}
\end{equation}

Following the analogy, we call the RHS of the continuity equation the \textit{flux across fixed wave number} $K$, or just the \textit{flux}, denoted by:

\begin{equation}\label{classicalflux}
^{\rho}P^4_K=-\int_0^Kdk\int_0^{\infty} dk_{123} d\Omega_{123} \rho_{k}\mathcal I^{12}_{3k'}
\end{equation}

The flux \eqref{classicalflux} has served as one of the effective metrics of conserved quantity transfer in wave turbulence, and has been used successfully as a theoretical test for numerical experiments involving wave turbulence \parencite{lvov2015formation,hrabski2022properties,newell2011wave,zhu2023direct,falcon2022experiments,shavit2024sign}.

{This notion of flux is constructed around the classical concept of the same name; interpreting the nonlocal flow of a conserved quantity through the wave number space to the flow through the surface of a sphere of radius $K$. However, by its construction the flux \eqref{classicalflux} is explicitly nonlocal, taking into account the entirety of interactions within the sphere $k<K$ while implicitly employing the interpretation of classical (local) flux through a surface. We will now show that our formalism \eqref{fourwaveflux} is compatible with the classical form \eqref{classicalflux}. We will also note the new formula expresses \eqref{classicalflux} in a way which demonstrates its nature as a nonlocal transfer of energy between sets, rather than a local flux through a surface.\\ 

To show the flux \eqref{classicalflux} is captured naturally by the transfer rate formula, we first multiply the integrand of \eqref{classicalflux} by the unity factor $\frac{\rho_1+\rho_2}{\rho_1+\rho_2}$, split the integral and apply a permutation of the indices  $1\leftrightarrow 2$, to obtain the form 

\begin{align}
&\begin{aligned}
^{\rho}P^4_K=-2\int_0^Kdk\rho_{k}&\Bigg[\int_0^Kdk_1\int_0^{\infty} dk_{23}d\Omega_{123}\frac{\rho_1\rho_\bk}{\rho_1+\rho_2}\mathcal I^{12}_{3k}\\
&+\int_K^\infty dk_1\int_0^{\infty} dk_{23} d\Omega_{123}\frac{\rho_1\rho_\bk}{\rho_1+\rho_2}\mathcal I^{12}_{3k}\Bigg]
\end{aligned}\\ 
&\quad \quad =\fp{\rho}^4_{(0,K)\to (0,K)}+\fp{\rho}^4_{(0,K)\to (K,\infty)}
\end{align}

Thus we have rewritten the flux \eqref{classicalflux} as a sum of transfers relayed by formula \eqref{fourwaveflux}. Let us focus on the transfer $\fp{\rho}_{(0,K)\to (0,K)}$ from a set to itself, which we term a \textbf{self interaction} \parencite{dematteis2023structure}. Due to the conservation of action and energy, we expect this quantity to be zero. Indeed, for some general set $A$, by the corresponding symmetries in the integral and the interaction terms we have for the general (not necessarily isotropic):

\begin{equation}
\begin{aligned}
\fp{\rho}_{A\to A}&=-2 \int_A d\bk d\bk_1\int d\bk_{23}\frac{\rho_\bk\rho_1}{\rho_1+\rho_2}\mathcal I^{12}_{3\bk}\\
&=- \int_A d\bk d\bk_1 \int d\bk_{23}\frac{\rho_1\rho_\bk(\rho_3+\rho_\bk-\rho_1-\rho_2)}{(\rho_1+\rho_2)(\rho_3+\rho_\bk)}\mathcal I^{12}_{3\bk}=0\\
\end{aligned}
\end{equation}

The final equality follows from the 
property of $\rho_\bk$ \eqref{resonantrho}, namely $\rho_1+\rho_2-\rho_3-\rho_\bk=0$.\\

Hence, the traditional equation for the fluxes in the scale invariant system \parencite{ZLF}, of flux across a fixed wave number $K$ \eqref{classicalflux}, is  expressed  as the transfer rate between sets $(0,K)$ and $(K,\infty)$ in the isotropic domain: 

\begin{equation}\label{FluxDefFirstTime}
^{\rho}P^4_{K}={^{\rho}P^4_{(0,K)\to (K,\infty)}}\,.
\end{equation}

The equation \eqref{FluxDefFirstTime} recasts the classical flux in a form which makes its nature clear; we see that the flux \eqref{classicalflux} can be equivalently described as capturing all of the conserved quantity transported from within a sphere of radius $K$ to its complement. The boundary of the sphere in this interpretation serves not as the location of flux, but of the boundary between regions of interaction. The RHS of the conservation form \eqref{ConservationForm} then captures the change in transfer as the role of wave numbers change with the radius $K$: adjusting $K$ denotes varying the regions of the wave number space which are ``sending" and ``receiving" in the flux calculation.  Further, the transfer rate formula representation \eqref{fourwaveflux} is a significant expansion to the flux: we are now free to express the dynamics of the wave kinetic equation in terms of conserved quantity transfer rates for systems which are fundamentally multidimensional without adhering to the assumptions surrounding isotropy.

The advantage of the \eqref{fourwaveflux} is that it does not require the isotropy
of the spectrum, and is applicable to anisotropic spectra as well. Furthermore, the calculation of fluxes at the Kolmogorov-Zakharov power law spectra below does not involve  L'Hôpital's rule to regularize an indeterminate limit \parencite{ZLF}.

Let us now consider the flux \eqref{FluxDefFirstTime} in the setting of stationary solutions. Written explicitly, the flux across fixed wave number $K$ for the isotropic four wave kinetic equation, with $\omega_\bk=\omega_k$ and $n_\bk= n_k$ and $S^{d-1}$ is the unit sphere, becomes \parencite{NazBook,ZLF}

\begin{equation}\label{IsotropicFlux}
\begin{aligned}
^{\rho}P^4_K=-2\int_0^{K}dk\:\rho_k \int_{K}^{\infty}d&k_1\int_0^{\infty} dk_{23}W^{\rho}(k_1,k_2)U^{12}_{3k}I^{12}_{3k}\delta(\omega^{12}_{3k})\\
U^{12}_{3k}=(k_1k_2k_3)^{d-1}\int_{S^{d-1}}dS^{d-1}\:|V^{12}_{3\textbf{k}}|^2&\delta(\bk+\bk_3-\bk_1-\bk_2);\quad W^{\rho}(k_1,k_2)=\frac{\rho_1}{\rho_1+\rho_2}
\end{aligned}
\end{equation}

We call $^\rho P^4_K$  the ``flux of conserved quantity $\rho_kn_k$ through/ across the wave number $K$.'' We make the following homogeneity assumptions for the existence of stationary solutions \parencite{ZLF,NazBook,galtier2022physics}

\begin{equation}\omega_k=k^{\alpha},\quad \rho_k=k^{\gamma},\quad V(a\bk_1,a\bk_2;a\bk_3,a\bk)=a^\beta V^{12}_{3\bk}\end{equation}

It follows then that the integrand of our flux is spatially homogeneous, where $U^{12}_{3k}$ has the homogeneity coefficient $2\beta+3d-4$, and the homogeneity coefficient of $W^{\rho}(\omega_1,\omega_2)I^{12}_{3k}\delta(\omega^{12}_{3k})$ is $3\nu-\alpha$, assuming $n_k=k^{\nu}$. If we rescale $\bk_i\to K\bk_i$, the flux across wave number $K$ can be written 

\begin{align}\label{generalscale}
^{\rho}P^4_K&=\left(K^{3d+2\beta+3\nu-\alpha +\gamma}\right) {^{\rho}P^4_{1}}
\end{align}
Then the flux is constant through wave number space (i.e. independent of $K$) if we choose the power law
\begin{align}
\nu=-\frac{1}{3}\left(2\beta-3d-\alpha+\gamma\right)\label{generalpower}
\end{align}

We have arrived at the celebrated Kolmogorov-Zakharov cascade solutions \parencite{Zakharov1967capillary,ZLF}, which by \eqref{ConservationForm} represent stationary solutions to $\dot n_k = 0$. We note this is not the first time this scaling property has been observed \parencite{NazBook}; however, in restricting the domain of the collision integral we render the integrand of the flux non-zero outside of pathological cases [citation 1-d surface grav waves]. Thus we find the natural conceptual ordering to be stationarity as a result of scale invariance in our framing of the wave kinetic equation.\\

Henceforth, we will denote the Kolmogorov-Zakharov powers $\nu_{\rho}$, where $\rho$ indicates the associated quadratic invariant weight $k^{\gamma}$. The spectra are named for their associated stationary solution of the wave kinetic equation: 

\begin{equation}
\begin{aligned}
\text{Constant-action-flux (``action'') Kolmogorov-Zakharov spectrum: }\nu_1 = -\frac{1}{3}\left(2\beta-3d-\alpha\right)\\
\text{Constant-energy-flux (``energy'') Kolmogorov-Zakharov spectrum: }\nu_{\omega} = -\frac{1}{3}\left(2\beta-3d\right)
\end{aligned}
\end{equation}

We see two properties emerge from the transfer formula at Kolmogorov-Zakharov solutions which will prove to have great utility in the latter sections. The first is a scaling property which follows directly from \eqref{generalscale} and the definition of the Kolmogorov-Zakharov solution; for any positive number $r$,
\begin{align}
^{\rho}P^4_{rA\to rB}=\:^{\rho}P^4_{A\to B}&\text{ for } n_k^{\nu_{\rho}}\label{ScaleInvar}\\
^{\omega}P^4_{rA\to rB}=(r^{\alpha}){^{\omega}P^{4}_{A\to B}}&\text{ for } n_k^{\nu_1}\label{scalevarianceenac}\\
^1P^4_{rA\to rB}=(r^{-\alpha}){^{\omega}P^{4}_{A\to B}}&\text{ for }n_k^{\nu_{\omega}}\label{scalevarianceacen}
\end{align}

These equations serve as an extension of the scale invariance property of flux to all transfer calculations. We recall that the ``action'' and ``energy'' Kolmogorov-Zakharov spectra were derived as the scale invariant solutions to the action and energy transfer formulas respectively. We note mismatching the two, calculating energy transfer for the action Kolmogorov-Zakharov spectrum and vice-versa, by \eqref{generalscale} will lead to the non-scale invariant forms (\ref{scalevarianceenac}) and (\ref{scalevarianceacen}). Scale invariance \eqref{ScaleInvar} only holds when the type of transfer matches the respective Kolmogorov-Zakharov solution. \\

The second property follows from the fundamental properties of the transfer formula \eqref{prop1leftsplit} and \eqref{prop2rightsplit}, combined with scale invariance \eqref{ScaleInvar}. Let $K_1<K_2<K_3$, and $A=(0,K_1)$ $B=(K_1,K_2)$ and $C=(K_2,\infty)$. Then at the $\rho$-Kolmogorov-Zakharov solution $n_k^{\nu_{\rho}}$,

\begin{equation}
\begin{split}
^{\rho}P^4_{A\to B}+{^{\rho}P^4_{A\to C}}={^{\rho}P^4_{(0,K_1)\to(K_1,\infty)}}\\
^{\rho}P^4_{A\to C}+{^{\rho}P^4_{B\to C}}={^{\rho}P^4_{(0,K_2)\to(K_2,\infty)}}
\end{split}
\end{equation}

We note by \eqref{ScaleInvar} these two sums are equal, so follows the physically intuitive property

\begin{equation}\label{equalinandout}
{^{\rho}P^4_{(0,K_1)\to (K_1,K_2)}}={^{\rho}P^4_{(K_1,K_2)\to (K_2,\infty)}}
\end{equation}

For the $\rho$ transfer associated to the $n^{\nu_\rho}_k$ Kolmogorov-Zakharov solution, the transfer of conserved quantity entering $(K_1,K_2)$ from the left is the same as the quantity leaving $(K_1,K_2)$ to the right.

\section{Spectral Transfers in Three, Five and Six Wave Kinetic Equations}\label{Theory_others}

The formalism presented in the
    previous section can  be generalized to the three-, five- and six-wave interacting systems. This is achieved in the present section.\\

\subsection{Three Wave Systems}\label{Section3wave}
\newcommand{\Jdense}[2]{{\dot{J}}_{#1}^{#2}}
\newcommand{\evalD}[1]{\big|_{\Delta(#1)}}

Three wave interaction systems model many important phenomenon including surface capillary waves\parencite{Zakharov1967capillary,pan2014direct}, oceanic internal waves, \parencite{lvov2001internal,dematteis2021downscale,dematteis2022origins,wu2023energy,labarre2024kinetics} and Rossby waves \parencite{NazRossbySurvey_2015,BALK1991_Rossby}. The nonlocal transfer of quadratic energy in three-wave systems was
introduced in \parencite{dematteis2023structure}. Using the formalism
established in this paper, section \ref{Theory_fourwave} above, we may re-represent this formulation to
study the transfer of additional invariant quantities. The associated three wave kinetic equation is \parencite{ZLF,dematteis2023structure,galtier2022physics,NazBook}

\begin{equation}
\begin{aligned}\label{eq:2}
\dot{n}_\textbf{k}= \int d\textbf{k}_{12}\calj^{\bk}_{12}\,,\qquad& \text{where}\\
\quad \calj^{\bk}_{12}=\calr^{\bk}_{12}-\calr^{1}_{\bk2}-\calr^{2}_{\bk1} 
,\qquad& \calr^{\bk}_{12}=4\pi|V^{\bk}_{12}|^2I^{\bk}_{12}\Delta^{\bk}_{12}\,,\\ I^{\bk}_{12} =n_1n_2n_{\bk}\left(\frac{1}{n_{\bk}}-\frac{1}{n_1}-\frac{1}{n_2}\right)\,,\qquad 
\Delta^{\bk}_{12}=&\delta(\bk-\bk_1-\bk_2)\delta(\omega_{\bk}-\omega_1-\omega_2)\,.
\end{aligned}
\end{equation}

The three wave kinetic equation \eqref{eq:2} relays its information quite differently than the four wave kinetic equation. We have two unique (up to symmetry) interaction terms, $\calr^{\bk}_{12}$ and $\calr^1_{\bk 2}$ each of which generates a resonant manifold that cannot be simultaneously satisfied by a fixed resonant trio $\bk_1,\:\bk_2,\:\bk$. This suggests that \eqref{eq:2} contains \textit{two} unique categories of scattering relationship suggested by the pairing of indices in the form $\bk \to (\bk_1,\bk_2)$ and $\bk_1\to (\bk,\bk_2)$. That is, the complete interactions of the ``active'' wave $\bk$ must include cases where $\bk$ scatters into two other waves, and where $\bk$ is the result of such a scattering. We will see this intuition reflected in the detailed balance below.

We now consider a conserved density $\eta_\bk n_\bk$ such that when the resonance condition $\Delta^{l}_{mn}$ is satisfied (where $l,\:m,\:n$ are d-dimensional vectors), i.e.

\begin{equation}\label{ResonantManifoldThreeWave}
\bk_l=\bk_m+\bk_n\,,\quad \omega_l=\omega_m+\omega_n\,,
\end{equation}

we have that

\begin{equation}
    \eta_{\textbf{l}}=\eta_{\textbf{m}}+\eta_{\textbf{n}}
\end{equation}

We introduce notation and properties: for some resonant trio $(\bk,\bk_1,\bk_2)$, let $\Delta(i)=\Delta^i_{(\cdot)(\cdot)}$ and $\calj^\bk_{12}\big|_{\Delta(i)}$ denote the integrand evaluated on the resonant manifold $\Delta(i)$. We define $\dot J^{12}_{\bk}$ to be the {detailed} interaction rate with respect $\bk$ and the resonant waves $\bk_1$, $\bk_2$, by 

\begin{align}{\dot{J}}^{12}_{\bk}\big|_{\Delta(i)}&=\eta_\bk\calj^{\bk}_{12}\big|_{\Delta(i)}\\
    \frac{1}{\eta_n}{\dot{J}}^{lm}_{n}\big|_{\Delta(l)}=-\frac{1}{\eta_l}{\dot{J}}^{mn}_{l}\big|_{\Delta(l)};&\quad
    \frac{1}{\eta_n}{\dot{J}}^{lm}_{n}\big|_{\Delta(l)}=\frac{1}{\eta_m}{\dot{J}}^{ln}_{m}\big|_{\Delta(l)}\label{Jequiv}
\end{align}

Then, ${\dot{J}}^{12}_{\bk}\big|_{\Delta(i)}$ is the detailed spectral interaction rate of wave $\bk$ with $\bk_1,\:\bk_2$ under the interaction type determined by index $i$. Thus, we have a detailed balance relation \parencite{dematteis2023structure, kraichnan1959structure}

\begin{equation}
\label{3WaveBalance}{\dot{J}}^{12}_{\bk}+{\dot{J}}^{2\bk}_{1}+{\dot{J}}^{1\bk}_{2}=0
\end{equation}

While the detailed balance is expressed over the entire collision integrand term, detailed balance can only be expressed in terms of a fixed trio of wave numbers as it handles individual interactions. Hence each resonant condition $\Delta(i)$ for $i=1,\:2,\:\bk$ will produce a distinct detailed balance relationship. Hence ~\eqref{3WaveBalance} will give rise to a detailed balance system of the form:

\begin{align}
    &\Jdense{\bk}{12}\evalD{\bk}+\Jdense{1}{1\bk}\evalD{\bk}+\Jdense{2}{1\bk}\evalD{\bk} = 0&
    \Jdense{\bk}{12}\evalD{\bk} = -\Jdense{1}{2\bk}\evalD{\bk}-\Jdense{2}{1\bk}\evalD{\bk}\\ \label{DetaledBalance3Start}
    &\Jdense{\bk}{12}\evalD{1}+\Jdense{1}{2\bk}\evalD{1}+\Jdense{2}{1\bk}\evalD{1} = 0
    \implies &\Jdense{1}{2\bk}\evalD{1}=-\Jdense{\bk}{12}\evalD{1}-\Jdense{2}{1\bk}\evalD{1} \\
    &\Jdense{\bk}{12}\evalD{2}+\Jdense{1}{2\bk}\evalD{2}+\Jdense{2}{1\bk}\evalD{2} = 0&\Jdense{2}{1\bk}\evalD{2}=-\Jdense{\bk}{12}\evalD{2}-\Jdense{1}{2\bk}\evalD{2}\label{DetailedBalance3End}
\end{align}

Similar to the four wave formalism above, we interpret each detailed
balance relationship as a scattering relationship, with
shared signs corresponding to being on the ``output'' side of the diagram. Thus to extract $\bk$'s interaction rate with respect to the wave(s) which is (are) scattering into it, \eqref{DetaledBalance3Start}-\eqref{DetailedBalance3End} is rewritten to group the interaction contributions by sign via \eqref{Jequiv}, and then use \eqref{Jequiv} to combine the ``output" terms which share a sign with $\Jdense{\bk}{12}$. We can then write $\calj^\bk_{12}$ over each of the resonance types to obtain the three wave transfer formula:

\begin{equation}\label{threewavetransfer}
\begin{aligned}
^{\eta}P^3_{A\to B}=\int_Ad\bk \int d\bk_{12}\Bigg[\chi_{B}(\bk_1)\frac{\eta_\bk}{\eta_1}\Jdense{1}{2\bk}\evalD{1}+\chi_{B}(\bk_2)\frac{\eta_\bk}{\eta_2}\Jdense{2}{1\bk}\evalD{2}+&\chi_{B}(\bk_1)\Jdense{1}{2\bk}\evalD{\bk}\\+&\chi_{B}(\bk_2)\Jdense{2}{1\bk}\evalD{\bk}\Bigg]
\end{aligned}
\end{equation}

This can also be expressed as
\begin{equation}\label{3wavetransfer}
^{\eta}P^3_{A\to B}=-2\int_Ad\bk\int_Bd\bk_1\int d\bk_2\:\eta_\bk\left[\calr^2_{1\bk}+\frac{\eta_1}{\eta_1+\eta_2}\calr^{\bk}_{12}\right]
\end{equation}

It was shown in the specific case of energy \parencite{dematteis2023structure} that, as the four wave system, the transfer formula recapitulates the isotropic flux \eqref{classicalflux} across a fixed wave number $K$ when $A=(0,K),\:B=(K,\infty)$ due to the self interactions being 0 as in the four wave case. The argument for the general transfer formula is equivalent and so will not be repeated.\\

The relevance of the extended formula can be seen in an application to Rossby waves studied via the  Charney–Hasegawa–Mima equation \parencite{NazRossbySurvey_2015, Hasegawa1, Hasegawa2}. Consider $\bk=(k_x,k_y)$; In addition to the two standard invariants momentum and energy given by $\eta_\bk = \bk$ and $\eta_\bk = \omega_\bk$ respectively, the equation has a unique invariant for a three wave system, zonostrophy given by \parencite{balk1990nonlocal,BALK1991_Rossby} $\eta_\bk=\arctan\left(\frac{k_y+\sqrt{3}k_x}{k^2}\right)-\arctan\left(\frac{k_y-\sqrt{3}k_x}{k^2}\right)$. { Note that the momentum component $k_x$ is constrained to be positive, and happens to be the collision invariant relative to enstrophy, which is therefore a supplementary sign-definite conserve quantity of the system.} Pairing the transfer equation \eqref{3wavetransfer} with the work presented in \parencite{NazRossbySurvey_2015}, for $A,B\subset (k_x,k_y)$, $k_x>0$, we may obtain expressions for the transfer of energy, enstrophy and zonostrophy, 
\begin{align}
^{\eta}P^3_{A\to B}=-2\int_Ad\bk\int_Bd\bk_1\int_{k_{2,x}>0}d\bk_2\:\eta_\bk\left[\calr^2_{1\bk}+\frac{\eta_1}{\eta_1+\eta_2}\calr^{\bk}_{12}\right]\\
\omega_k=-\frac{k_x}{k^2+F}\,,\quad
V^{\bk}_{12}=\text{sgn}(k_x)\sqrt{\beta k_{1,x}k_{2,x}k_x}\left(\frac{k_{1,y}}{k_1^2+F}+\frac{k_{2,y}}{k_2^2+F}-\frac{k_y}{k^2+F}\right)\\
\eta_\bk = k_x:\:\text{Enstrophy}\quad \eta_\bk = \omega_{k,x}:\:\text{Energy}\\
\eta_\bk=\arctan\left(\frac{k_y+\sqrt{3}k_x}{k^2}\right)-\arctan\left(\frac{k_y-\sqrt{3}k_x}{k^2}\right):\:\text{Zonostrophy}
\end{align}
In this particular example, the $\omega_k$ is not positive definite, but negative definite. Our formalism is totally applicable here and the study of Rossby waves 
 via the Charney–Hasegawa–Mima equations proves relevant to this work because the enstrophy and zonostrophy flow betray very nonlocal transfers \parencite{NazRossbySurvey_2015}. The ability to study the direct transfer of these collision invariants directly between different regions through \eqref{3wavetransfer} may prove a unique and useful tool.

\subsection{Five Wave Systems - unidirectional water waves on top of deep ideal fluid}


A five wave kinetic equation was derived for gravity waves on deep water in \parencite{Krasitskyfivewave,DYACHENKOLVOVZAKHAROV1995FiveWave}.

\newcommand{\calf}{\mathcal{F}}

\begin{align}\label{fivewave}
\rho_{\bk} \dot n_{\bk}=\int d\bk_{1234}\rho_{\bk}&\left(\frac{1}{3}f^{123}_{4 \bk}-\frac{1}{2}f^{12\bk}_{34}\right)\\
f^{123}_{\bk 4}=\pi|V^{123}_{\bk 4}|^2\delta^{123}_{4\bk}\delta(\omega^{123}_{4\bk})n_1n_2n_3n_4&n_{\bk}\left[\frac{1}{n_4}+\frac{1}{n_{\bk}}-\frac{1}{n_1}-\frac{1}{n_2}-\frac{1}{n_3}\right]
\end{align}

The symmetries of $V^{123}_{\bk4}$ occur by permutations of within the indices $(1,2,3)$ and $(4,\bk)$, with no symmetries between the two tuples. Following from these, the symmetries in the individual interaction terms are of the form $f^{123}_{4\bk}=f^{123}_{\bk4}=f^{213}_{4\bk}=f^{231}_{4\bk}$. Based upon our intuition from the three wave case, we can see that the collision integrand of \eqref{fivewave} will generate at least two unique interaction types, in this case $2\to 5$ and $5\to 2$ wave scattering relationships with distinct resonant manifolds. However, to obtain the scattering relationships in terms of sign, we will have to express detailed balance in terms of the entire collision integrand $\calf^{1234}_{\bk}$. The detailed balance relationship emerges when we use the symmetries of the integral to form all combinations of the wave numbers $\bk_1,\:\bk_2,\:\bk_3,\:\bk_4,$ with $\bk$ that cannot otherwise be achieved by the symmetries present in the interaction terms:

\begin{equation}\label{fivewaveinteractionform}
\begin{aligned}
&\int d\bk_{1234}\rho_{\bk}\left(\frac{1}{3}f^{123}_{4 \bk }-\frac{1}{2}f^{12\bk}_{34}\right)\\
&=\int d\bk_{1234}\frac{\rho_{\bk}}{12}\left(f^{123}_{4\bk}+f^{124}_{3\bk}+f^{134}_{2\bk}+f^{234}_{1\bk}-f^{12\bk}_{34}-f^{13\bk}_{24}-f^{14\bk}_{32}-f^{23\bk}_{14}-f^{24\bk}_{13}-f^{34\bk}_{12}\right)\\
&=\int d\bk_{1234}\calf^{1234}_{\bk}
\end{aligned}
\end{equation}

In this form, we see that $\calf^{1234}_{\bk}$ satisfies detailed conservation when $\rho_\bk n_\bk$ satisfies the resonance condition $\rho_\bk+\rho_4=\rho_1+\rho_2+\rho_3$ for a fixed quintet defined by $\Delta^{123}_{4\bk}=\delta^{123}_{4\bk}\delta(\omega^{123}_{4\bk})$,

\begin{equation}
\calf^{1234}_{\bk}+\calf^{\bk234}_{1}+\calf^{1\bk34}_{2}+\calf^{12\bk4}_{3}+\calf^{123\bk}_{4}=0
\end{equation}

We see balance occur when we count the symmetric terms uniquely, balancing the weights $\frac{1}{2}$ and $\frac{1}{3}$ with the degrees of freedom up to symmetry of each of the two interaction types.

In this case, the detailed balance form of the integrand induces ten unique resonant manifolds; however we note that there are only two core forms produced: $\Delta^{(\cdot)(\cdot)(\cdot)}_{(\cdot)\bk}$ and $\Delta^{(\cdot)(\cdot)\bk}_{(\cdot)(\cdot)}$. We will present only one representative of each type for brevity. We proceed following the program of the three wave case presented above in the section \ref{Section3wave}. Namely, we fix a resonant manifold and use the detailed balance to collect interacting terms of the same sign on each side.\\

\begin{align}
&\calf^{1234}_{\bk}\Big|_{\Delta^{123}_{4\bk}}+\calf^{123\bk}_{4}\Big|_{\Delta^{123}_{4\bk}}=-\calf^{\bk234}_{1}\Big|_{\Delta^{123}_{4\bk}}-\calf^{1\bk34}_{2}\Big|_{\Delta^{123}_{4\bk}}-\calf^{12\bk4}_{3}\Big|_{\Delta^{123}_{4\bk}}\\
&\calf^{1234}_{\bk}\Big|_{\Delta^{12\bk}_{34}}+\calf^{\bk234}_{1}\Big|_{\Delta^{12\bk}_{34}}+\calf^{1\bk34}_{2}\Big|_{\Delta^{12\bk}_{34}}=-\calf^{12\bk4}_{3}\Big|_{\Delta^{12\bk}_{34}}-\calf^{124\bk}_{4}\Big|_{\Delta^{12\bk}_{34}}\\
&\hspace{5cm}\downarrow\nonumber\\
&\calf^{1234}_{\bk}\Big|_{\Delta^{123}_{4\bk}}=\frac{\rho_{\bk}}{\rho_4+\rho_{\bk}}\left(-\calf^{\bk234}_{1}\Big|_{\Delta^{123}_{4\bk}}-\calf^{1\bk34}_{2}\Big|_{\Delta^{123}_{4\bk}}-\calf^{12\bk4}_{3}\Big|_{\Delta^{123}_{4\bk}}\right)\\
&\calf^{1234}_{\bk}\Big|_{\Delta^{12\bk}_{34}}=\frac{\rho_{\bk}}{\rho_1+\rho_2+\rho_{\bk}}\left(-\calf^{12\bk4}_{3}\Big|_{\Delta^{12\bk}_{34}}-\calf^{124\bk}_{4}\Big|_{\Delta^{12\bk}_{34}}\right)
\end{align}

These two cases are sufficient, as the remaining manifolds are formed by the permutations up to symmetry of $(1,2,3,4)$ in $\Delta^{123}_{4\bk}$, for which we have four in total and in $\Delta^{12\bk}_{34}$ for which we have six in total.
\begin{equation}
\begin{split}
^{\rho}P^5_{A\to B}=-\int_Ad\bk \rho_\bk \int d \textbf{k}_{1234}
\left(\frac{\chi_B(\textbf{k}_1)\rho_1f^{123}_{4\bk}}{\rho_1+\rho_2+\rho_3}-\frac{\chi_B(\textbf{k}_3)\rho_3f^{12\bk}_{34}}{\rho_3+\rho_4}\right)
\end{split}
\end{equation}

The coefficients $1/2$ and $1/3$ normalize $f^{123}_{4\bk}$ and $f^{12\bk}_{34}$ based upon the number of symmetries in the waves which $\bk$ scatters into. We see that when these symmetries were made explicit in the individual detailed balance relationships, the coefficients are balanced by the number of scattering cases for $\bk$ in each case written in \eqref{fivewaveinteractionform}, and we see them vanish once the integral symmetry is applied, with all combinatorial cases of wave scattering being captured in the transfer formula.

\subsection{Six Wave Systems - three-to-three interactions}

It was shown in \parencite{Onorato2015PNAS} that
the celebrated alpha-FPU problem is goverened by six wave interactions
\parencite{Onorato2015PNAS}. And there are further examples in optical
turbulence \parencite{Bortolozzo09sixwaveoptical}. We therefore extend
our formalism to the six wave kinetic equation.  Here we consider only
the process of three waves scattering into three other waves, or the
three-to-three interactions. The generic three-wave-to-three-wave scattering six wave Hamiltonian
equation is given by \parencite{NazBook}

\begin{equation}
\mathcal{H}=\int d\bk \left(\omega_\bk a_\bk a_\bk^* + \frac{1}{3}V^{\bk12}_{345}\delta(\bk^{\bk12}_{345})a_\bk^*a_1^*a_2^*a_3a_4a_5\right)
\end{equation}

Where $V^{\bk12}_{345}$ has permutation symmetries within the tuples $(\bk,1,2)$ and $(3,4,5)$ as well as the reality condition $V^{\bk12}_{345}=(V_{\bk12}^{345})^*$. The associated wave kinetic equation is \parencite{NazBook,Bortolozzo09sixwaveoptical}
\begin{align}
\dot n_\bk =& \int d\bk_{12345}\cali^{12\bk}_{345}\\
\cali^{12\bk}_{345}=12\pi |V^{12\bk}_{345}|^2\Delta^{12\bk}_{345}&n_1n_2n_3n_4n_5n_\bk\left(\frac{1}{n_\bk}+\frac{1}{n_1}+\frac{1}{n_2}-\frac{1}{n_3}-\frac{1}{n_4}-\frac{1}{n_5}\right)
\end{align}

Where $\cali^{12\bk}_{345}=-\cali_{12\bk}^{345}$ has symmetries within $(12\bk)$ and $(345)$. Due to the presence of these symmetries, the detailed balance relationship can be written on a single resonant manifold.

This observation allows us to extend our previous consideration to the
more general $\rho_{\bk}$ satisfying
$\rho_1+\rho_2+\rho_\bk=\rho_3+\rho_4+\rho_5$ on
$\Delta^{12\bk}_{345}$,
\begin{equation}
\begin{split}
^{\rho}P^6_{A\to B}=-3\int_Ad\textbf{k}\:\rho_\bk \int d\textbf{k}_{12345}  \left(\frac{\rho_3\chi_B(\bk_3)}{\rho_3+\rho_4+\rho_{5}}\right)\cali^{12\bk}_{345}
\end{split}    
\end{equation}

We have observed detailed balance serves as a
backbone property shared by these wave kinetic equations.  The
detailed balance along with the interaction diagram interpretation of
interaction dynamics contains a wealth of information on the structure
of interaction. Indeed, in the three wave case, we extended the
formalism to handle {different conserved quantities other than energy}. 
Furthermore, in the five wave case we discovered that
symmetries in the structure of the collision integrand are not a
necessary feature to develop a detailed balance relationship. The
formalism presented in this manuscript appears to be flexible and
universal.

\section{Application: The Structure of Cascades in the Majda-McLaughlin-Tabak  Model \label{Numerics}}

\subsection{Majda-McLaughlin-Tabak model: setting up the stage}

   Perhaps the most intriguing four-wave system is the surface
      gravity waves~\parencite{Z68a}.  Yet, the surface gravity wave problem is too
      complex, and to test the wave turbulence a simpler system was
      needed. To provide a setting to study the fundamentals of four-wave interactions, A. Majda, D. McLaughlin and E.
      Tabak introduced a simpler phenomenological model (MMT) in 1997
      \parencite{MMT}. The Majda-McLaughlin-Tabak model is a
      classical example of a generic four-wave Hamiltonian
      \parencite{ZLF, NazBook, galtier2022physics} system which
      contains only the scattering of two waves into two other
      waves. The model can be set to have the linear dispersion
      relationship of surface gravity waves on top of deep ideal
      fluid. Remarkably, the Majda-McLauglin-Tabak model allows for
      the direct numerical modeling of the underlying equations of
      motion.\\

  The Majda-McLaughlin-Tabak model has been intensively investigated in the past three decades\parencite{CaiMMT1999,CAIMMT,ZAKHAROVPush1_DMMT,rumpfmmtpulse,onoratoMMT2016,hrabski2023verification,buhlerMMT2023,simonis2024time,simonis2024transition}.
  In the present
  manuscript, we will investigate the properties of the transfers of wave action and energy in the
  Majda-McLaughlin-Tabak model.

\subsection{Isotropic Transfer Formula for Majda-McLaughlin-Tabak model}

The Majda-McLaughlin-Tabak  Hamiltonian is given by \parencite{MMT}

\begin{equation}
\mathcal{H}=\int dk \omega_k|a_k|^2+\frac{\lambda}{2} \int d k_{123k}V^{12}_{3k} a_1a_2a^*_3a^*_k \delta^{12}_{3k}.
\end{equation}
Here, as before all the integrals are taken over all available space, the real 
line in this case. Furthermore, the notation
$\int dk_{123}$ implies the integrals over $k_1$, $k_2$ and $k_3$.

In the Majda-McLauglin-Tabak model the linear dispersion relationship
is given by
\begin{equation}\label{eq:MMTdisp}
\omega_k=|k|^{\alpha}    
\end{equation}
The interaction matrix element, describing the strength of interactions
between the four wave numbers is given by
\begin{equation}\label{eq:MMTme}
V^{12}_{3k}=|k_1k_2k_3k|^{-\beta/4}
\end{equation}
Following the original Majda-McLaughlin-Tabak paper \parencite{MMT} 
we choose $\alpha =\frac{1}{2}$ as in surface gravity waves on deep water \parencite{Z68a} (and because it the resonant manifold is easily expressed), and will, unless otherwise noted, choose
$\beta=\frac{1}{4}$ for all calculations.

The wave kinetic equation associated with this two-to-two wave
Hamiltonian is given by \eqref{KineticEqnFourWave}. 

The resonant manifold is given by quartets $k,\:k_1,\:k_2,\:k_3$ which satisfy the resonant conditions
\begin{align}
&k_1+k_2=k_3+k\label{UnaveragedMan a}\\
&|k_1|^{\alpha}+|k_2|^{\alpha}=|k_3|^{\alpha}+|k|^{\alpha}\label{UnaveragedMan b}
\end{align}
Since  $\alpha=\frac{1}{2}$, one of the wave numbers has the direction
opposite to the other three, \parencite{DYACHENKOLVOVZAKHAROV1995FiveWave}
, and the resonant manifold may be parametrized as

\begin{equation}\label{parameterization}
k_1=a(1+\zeta)^2,\quad k_3=a(1+\zeta)^2\zeta^2,\quad k_1=-a\zeta^2,\quad k_2=a(1+\zeta+\zeta^2)^2\,.
\end{equation}
Note note that this parameterization captures all \textit{nontrivial}
resonant quartets; the additional solutions to the resonant manifold
such as $k_1=k,\:k_2=k_3$ are trivial so that the collision integrand
is identically equal to zero. Furthermore, this parameterization
\eqref{parameterization} permits us to write a sign-averaged collision
integral, where we extract all possible cases of resonance in terms of
positive wave numbers. Thus, given $A=[a_1,a_2]$ and $B=[b_1,b_2]$
such that $a_i<b_i$, we may write a sign averaged four wave transfer
formula for the Majda-McLaughlin-Tabak model as
\begin{equation}
\begin{split}\label{angleavMMT}
^{\rho}P^4_{A\to B} = -8\pi \int_A dk\int_B dk_1\int_0^{\infty} dk_{23} \:\frac{\rho_1\rho_k}{\rho_1+\rho_2}|V^{12}_{3k}|^2 \delta(\omega^{12}_{3k})I^{12}_{3k}\cdot\\
\left[ \delta(|k_1|-|k_2|+|k_3|+|k|)+\delta(|k_1|-|k_2|-|k_3|-|k|)\right.\\
\left.+\delta(|k_1|+|k_2|+|k_3|-|k|)+\delta(|k_1|+|k_2|-|k_3|+|k|)\right]
\end{split}
\end{equation}

\paragraph{{Contextual Simplification of the Resonant Manifold}}

Consider the second and fourth resonance conditions
in \eqref{angleavMMT}. These, paired with the
Dirac Delta functions over the difference of the frequencies
$\delta(\omega^{12}_{3k})$ form the system
\begin{align}
\omega_1+\omega_2-\omega_3-\omega_k=0\label{resmanifold a}\\
\omega_2^{1/\alpha}+\omega_3^{1/\alpha}+\omega_k^{1/\alpha}=\omega_1^{1/\alpha}\label{resmanifold b}\\
\omega_k^{1/\alpha}+\omega_1^{1/\alpha}+\omega_2^{1/\alpha}=\omega_3^{1/\alpha}\label{resmanifold c}
\end{align}

\begin{figure}
    \hspace{0cm}
    \subfloat[]{\includegraphics[scale=.3]{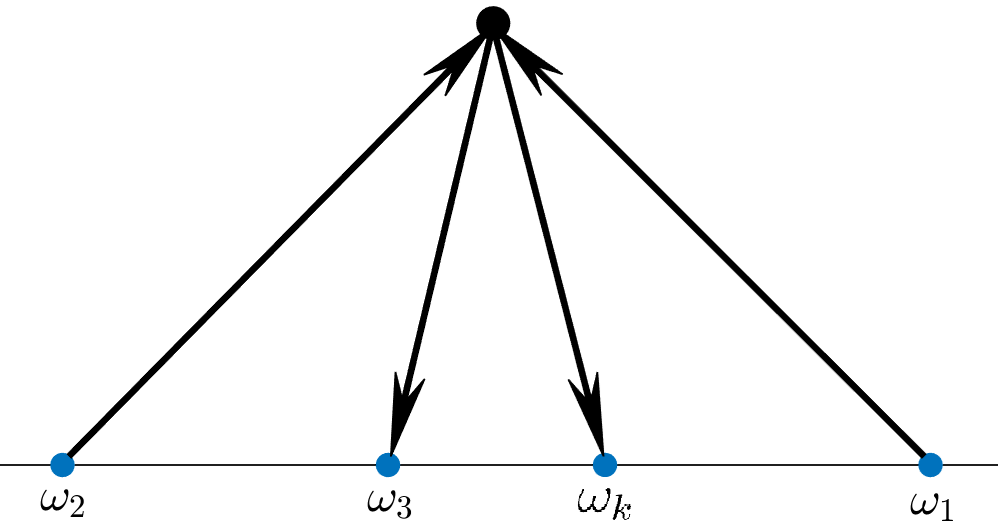} \label{rondo1}}
    \hspace{.15cm} \subfloat[]{\includegraphics[trim=0 .15cm 0 0,scale=.3]{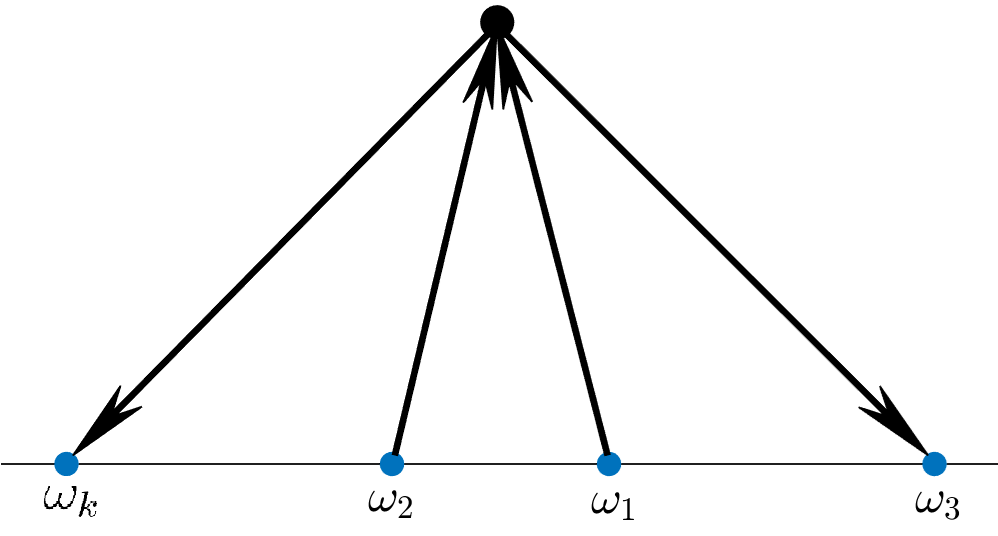} \label{rondo2}}\\

    \caption{A schematic representation of the valid solutions of resonant manifold when $\omega_k$ increases as a function of $k=|\bk|$. Figure (a) corresponds to a case which generate constituent integral $I^{[1]}$, and figure (b) corresponds to a case which generate constituent integral $I^{[2]}$. {Due to the  energy conservation, the sum of the two frequencies in the middle must equal the sum of the other two frequencies. This implies that in each individual quartet energy is always transferred simultaneously both forward and backward.}}\label{schematicresman}   

\end{figure}

Taking into account the symmetries of the resonant manifold, and using 
\eqref{resmanifold b} and \eqref{resmanifold c} we have, respectively

\begin{equation}\label{resmandecomp}
\omega_2<\omega_k<\omega_1\, , \quad
\omega_2<\omega_3<\omega_1\:; \ \ 
\quad
\omega_k<\omega_1 < \omega_3\, ,\quad 
\omega_k<\omega_2 < \omega_3\, .
\end{equation}

Where the first case follows from \eqref{resmanifold b} and the second from \eqref{resmanifold c}. By \eqref{resmandecomp}, we observe that the first and third resonance conditions in \eqref{angleavMMT} are not relevant to our formulation when assuming for all $a\in A,\:b\in B$ that $a<b$. {{Additionally, these two cases of resonant manifold construction are captured disjointly by the resonance conditions \eqref{resmanifold b}-\eqref{resmanifold c}, permitting us to split the transfer formula into a positive and negative constituent integral (Details provided in appendix) 
\begin{equation}\label{thing1}
\begin{split}
\fp{\rho}^4_{(a_1,a_2)\to (b_1,b_2)}=  I^{[1]}+I^{[2]};\quad I^{[1]}<0<I^{[2]}
\end{split}
\end{equation}
}}\\

With these simplifications, the MMT isotropic transfer formula is a numerically amenable expression. It is well known \parencite{ZLF,galtier2022physics,NazBook} that the four wave kinetic equation has four steady state solutions $n_k=|k|^x$ where (with the MMT $\beta$ parameter negative)

\begin{equation}
\begin{aligned}
x_{\text{particle equipartition}}=0 &\quad x_{\text{energy equipartition}}=-\alpha\\
x_{\text{energy flux}}=-\frac{1}{3}\left(3-2\beta\right)&\quad x_{\text{action flux}}=-\frac{1}{3}\left(3-2\beta-\alpha\right)
\end{aligned}
\end{equation}

For consistency, we fix the parameters $\alpha=.5$ and $\beta=.25$ in our calculations. In this case, the steady state solutions are given by

\begin{equation}
\begin{aligned}
x_{\text{particle equipartition}}=0 &\quad x_{\text{energy equipartition}}=-\frac{1}{2}\\
x_{\text{energy flux}}=-\frac{5}{6}&\quad x_{\text{action flux}}=-\frac{2}{3}
\end{aligned}
\end{equation}

These stationary spectra, which are the solutions of scale invariant flux, serve as the cornerstone of our investigation below.

\subsection{Global structure of fluxes}

\begin{figure}
    \hspace{-2.2cm}
    \includegraphics[scale=.65]{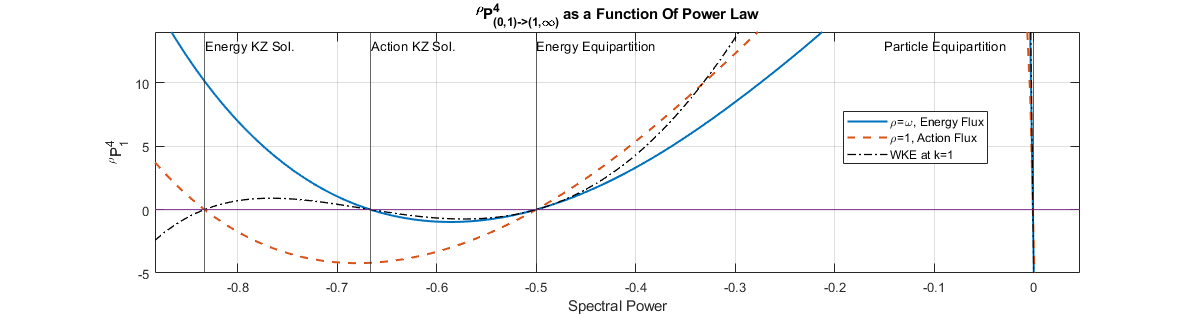}
    \caption{The flux across fixed wave number $K=1$, $^{\rho}P^4_1$ calculated at $n_k^{\nu}$ as a function of spectral power law. We see that the energy flux $^{\omega}P^4_1$ is positive at the energy Kolmogorov-Zakharov solution $n_k^{\nu_{\omega}}$ and 0 at $n_k^{\nu_1}$ corresponding to a forward cascade with 0 flux on the inverse cascade solution $n_k^{\nu_1}$. The opposing set of conditions hold for action flux. We additionally include the wave kinetic equation evaluated at $k=1$; this is 0 at both Kolmogorov-Zakharov solutions corresponding to the scale invariance of flux at stationary states. {{Note that all three curves are almost vertical and almost coincide at Spectral Power near zero.}}
    }
    \label{MatchFig}
\end{figure}

Our formalism allows us to calculate the amount of flux of
    spectral energy density or flux of wave action through the given
    wave number.  As a first step, we calculate numerically the
    flux of wave action and the flux of energy across the given wave
    number.  We choose the set $A$ as all wave numbers such that
    the modulus of the wave number is less than given wave number
    $k^*$, and set $B$ such that the modulus of wave numbers is larger
    than $k^*$.  Since both the linear dispersion relationship \eqref{eq:MMTdisp}  and the
    interaction matrix elements \eqref{eq:MMTme} are
    scale invariant, we choose $k^* =1$. We choose the wave action
    to be $n_k = |{\bf k}|^{x}$, and we calculate the flux of energy
    and the flux of wave action from set $A$ to set $B$ as a function
    of $x$. 
    The results are shown in Figure \ref{MatchFig}.
    We also show the value of the wave kinetic equation
    collision integral \eqref{KineticEqnFourWave} as a function of
    $x$.
    We observe the following:
    \begin{itemize}
    \item There are four zeros of the collision integral: 
 two zeros   corresponding to Kolmogorov Zakharov solutions
 of constant fluxes of energy and wave action, and two 
 equipartition solutions of energy and wave action;
 \item The flux of energy is zero on the constant wave action flux
 solution and on the equipartition of energy;
 \item The flux of wave action is zero on the constant energy flux and equipartition of wave action
 \item The flux of energy is positive on the constant energy flux,
 i.e. we observe {\it{direct}} cascade of energy;
 \item The flux of wave action is negative on the wave action flux, i.e. there is an {\it{inverse}} cascade of wave action;
 \end{itemize}

 {{While these statements are quite intuitive, our formalism
 provides a mathematical, albeit numerical proof of their 
 correctness.}} {{Arguments for the sign of the flux of the cascades have been presented in \parencite{ZLF,NazBook}, and the fact that the flux is 0 on counter-transfer types follows from stationary \eqref{ScaleInvar} and the properties \eqref{scalevarianceenac} and \eqref{scalevarianceacen}.}}

{{Our analysis focuses on the wave action and energy Kolmogorov-Zakharov solutions $n_k^{\nu_1}$ and $n_k^{\nu_{\omega}}$. The four calculated characterizations of these spectra in figure \ref{MatchFig} suggest four interpretations: each spectrum can be considered both a solution of constant flux or a solution of zero flux to the energy and action continuity equations}}

\begin{align}
\int_0^kds\: \dot n_s = {^1P^4_{(0,k)\to (k,\infty)}}\label{WKE_action}\\
\int_0^k ds\: \dot \varepsilon_s =\int_0^k ds\: \omega_s\dot n_s= {^{\omega}P^4_{(0,k)\to (k,\infty)}}\label{WKE_energy}
\end{align}

{{The continuity equations can be interpreted from the perspective of classical flux, as we have done up until now by treating $k$ as the sole relevant parameter. The transfer equation, however, permits us the luxury of studying precisely the structure of transfers underlying the stationary solutions to \eqref{WKE_action}-\eqref{WKE_energy} which are dependent on all resonant interactions. There are a multitude of possible decompositions of the fluxes on the RHS of the continuity equations which could answer various questions, such as the transfer rates from a forcing to dissipation region, the transfer rates from an inertial range into a dissipation region, or other such contextual queries. As we are working with pure theoretical Kolmogorov-Zakharov solutions we will, in this section, decompose the flux to study the overall structure of conserved quantity transfer using property \eqref{prop1leftsplit} as}}

\newcommand{\picscale}{.57}
\newcommand{\picscaletwo}{.57}

\subsubsection{Nonlocal transport cumulative distribution}\label{sec:4.4.1}

\begin{figure}
    \hspace{-.1cm}
    \subfloat[]{\includegraphics[scale=\picscaletwo]{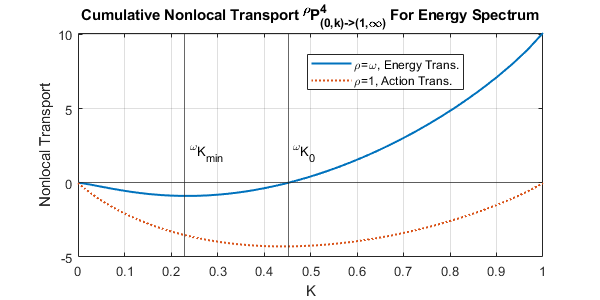} \label{NonlocalEnergyLeftRight}}
    \hspace{-1cm}
       \subfloat[]{\includegraphics[scale=\picscaletwo]{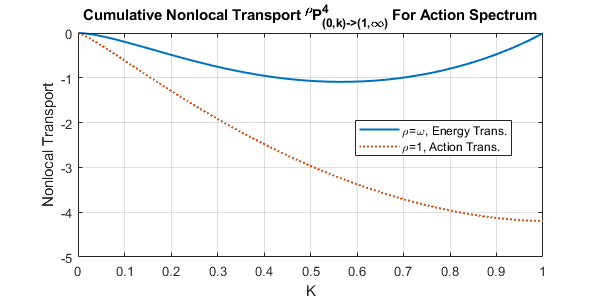} \label{NonlocalActionLeftRight}}\\
       \hspace{-.5cm}
    \subfloat[]{\includegraphics[scale=\picscale]{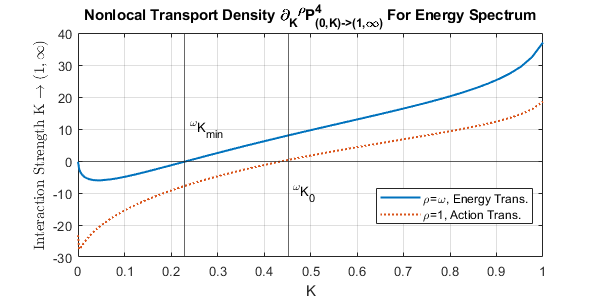} \label{NonlocalTransDerivEn}}
    \hspace{-1cm}
    \subfloat[]{\includegraphics[scale=\picscale]{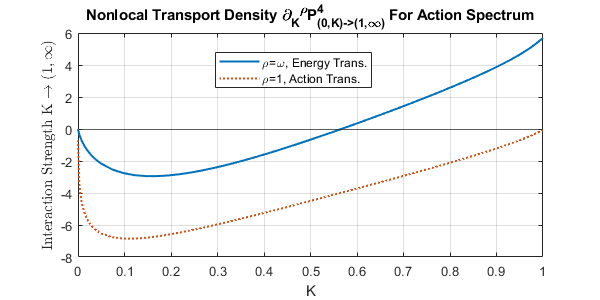} \label{NonlocalTransDerivAc}}   
    \caption{The four cases of the nonlocal transport curve calculated for energy and action Kolmogorov-Zakharov spectra. At $K=1$ both cases obtain the flux $^{\rho}P^4_{1}$. (a) We note the inverse energy transfer mechanism between when $K\in \left(0,{^{\omega}K_0}\right)$.
    (b) The nonlocal transports for the action cascade solution, this appears more nonlocal than energy. (c) Point to set transport for the energy spectrum (d) The same calculation as performed for a fixed action Komogorov-Zakharov spectrum.}\label{nonlocaltrans}
\end{figure}

\begin{equation}\label{decompositionofflux:1}
\int_0^k ds\: \rho_s\dot n_s = {^{\rho}P^4_{(0,k)\to (k,\infty)}}=\:{^{\rho}P^4_{(0,K)\to (k, \infty)}}+\:{^{\rho}P^4_{(K,k)\to (k,\infty)}};\quad 0<K<k
\end{equation}

{{We refer to $^{\rho}P^4_{(0,K)\to (k,\infty)}$ as the \textbf{nonlocal transport {cumulative distribution}} and $^{\rho}P^4_{(K,k)\to (k,\infty)}$ as the \textbf{local transport {cumulative distribution}}. We choose to study the transfer structure with respect to decomposition of the sending set $(0,k)$ as this leads to a compact domain for the choice of $K$ which is convenient for visualization. We note that using the scaling properties of the Kolmogorov-Zakharov solutions \eqref{ScaleInvar}-\eqref{scalevarianceacen} with the properties of the transfer formula \eqref{prop1leftsplit}-\eqref{prop2rightsplit} we may rewrite these fundamental calculations in other forms which decompose the receiving set $(k,\infty)$. We will thus study the structure of transfer underlying the stationary solutions to the continuity equations.}}\\

\subsection{Nonlocal Transport}

{\black{We now  investigate the relative importance of the nonlocal contributions to the flux. After making the choice $k=1$ in Eq.\eqref{decompositionofflux:1}, we now vary $K$ and study the nonlocal transport cumulative distributions of wave action ($^{1}P^4_{(0,K)\to (1,\infty)}$) and energy ($^{\omega}P^4_{(0,K)\to (1,\infty)}$) for the two Kolmogorov-Zakharov spectra of constant action flux $n_k^{\nu_1}$ and constant energy flux $n_k^{\nu_{\omega}}$.  The results are presented in Fig. 
\ref{nonlocaltrans}. }} First, notice that both cumulative transports of energy and action are zero 
for $K=0$ for both Kolmogorov-Zakharov solutions. This is due
to the fact that set $A$ is null if $K=0$.
{\black{Then, we observe the 
following:\\
{\bf Direct energy cascade solution $n_k^{\nu_{\omega}}$, Fig.~\ref{nonlocaltrans}(a)}
\begin{itemize}
\item The energy cumulative transport has a zero at $K=^\omega K_0\simeq0.45$. It is positive for $K>^\omega K_0$, indeed approaching the (positive) ``total'' flux of the direct energy cascade as $K\to1$ ({\it cf.} energy Kolmogorov-Zakharov solution in Fig. \ref{MatchFig}). However, it is negative for $K<^\omega K_0$, and we find this fact fascinating and counterintuitive. Within the direct energy cascade, there is a hidden inverse energy cascade that rules the more nonlocal transfers (left side of the plot) - whereas the direct transport is ruled by more local transfers (right side of the plot). This fact mirrors the observations of Cai, Majda, Mclaughlin and Tabak \parencite{CaiMMT1999} that there are mechanisms of energy return which occur nonlocally during an energy cascade of the MMT model.
\item We observe that there is a nonzero cumulative transport of wave action for $0<K<1$.
It is zero at both $K=0$ and $K=1$. 
For $K=0$ the set $A$ is null, and for $K=1$ there is no total wave action
flux for the direct energy cascade Zakharov-Kolmogorov solution. 
Yet, the nonzero wave action flux for $0<K<1$ indicates that there is an 
inverse cascade of wave action, which is balanced to zero as $K$ approaches $1$.
\end{itemize}
{\bf Inverse action cascade solution $n_k^{\nu_{1}}$, Fig.~\ref{nonlocaltrans}(b)}
\begin{itemize}
\item For the constant wave action flux solution there is also an effective energy
flux which balances out to zero as $K$ approaches $1$.
\item The entire action nonlocal transport is negative, starting from zero at $K=0$ and ending up at the value of the total action flux at $K=1$ ({\it cf.} action Kolmogorov-Zakharov solution in Fig. \ref{MatchFig}).
\end{itemize}
These observations allow us to delve into the ``mechanics" of the celebrated Kolmogorov-Zakharov solutions 
of the wave kinetic equation.
}}

\subsubsection{Nonlocal transport density distribution}

{{To further investigate  the dynamical structure of the cascades, we will now 
 focus on}} the distribution of transfers between individual wave numbers and intervals. 
 We  now consider the general transfer rate between sets $A=[a,a']$ and $B=[b,b']$ (taking $b>a'$ and $b'\to\infty$). Then by property \eqref{prop1leftsplit} we can express a difference quotient of the transfer rate as follows 

\begin{align}
\frac{{^{\rho}P^4_{A\to B}}}{a'-a}&=\frac{{^{\rho}P^4_{(0,a')\to B}}-{^{\rho}P^4_{(0,a)\to B}}}{a'-a}\nonumber\\
 \int_0^{\infty}dk\frac{\chi_{A}(k)}{a'-a}\int_1^{\infty}dk_1\int dk_{23}\rho_k{^{\rho}W(k_1,k_2)}\cali^{12}_{3k}&=\frac{{^{\rho}P^4_{(0,a')\to B}}-{^{\rho}P^4_{(0,a)\to B}}}{a'-a}\label{derivextraresonance}\\ 
\xrightarrow{a' \to a}\int_0^{\infty} dk \int_0^1dk_1\int_{0}^{\infty} dk_{23}\delta(k-a)\rho_k{^{\rho}W(k_1,k_2)}\cali^{12}_{3k}&=\partial_a {^{\rho}P^4_{(0,a)\to B}}\nonumber
\end{align}
Comparing the beginning and the end of \eqref{derivextraresonance} we obtain
\begin{align}
&^{\rho}P^4_{a\to B}=\partial_a{^{\rho}P^4_{(0,a)\to B}}\label{wnatoB}
\end{align}

We see that the derivative of the nonlocal transport curve between sets $[0,a]$ and $B$ is equivalent to introducing a restriction on the transfer formula in the form $\delta(k-a)$. Therefore  the derivative represents the infinitesimal transfer rate between set $B$ and the individual wave number $a$. We call this function the {\bf nonlocal transport density}. Looking back to the scattering diagram (Fig. \ref{fig:ScatteringDiagram}), this is the net interaction rate between all resonant waves $k_1\in B$ and the individual wave $a$.

The nonlocal transport density of the two Kolmogorov-Zakharov solutions are shown in Fig.~\ref{nonlocaltrans}(c)-(d).
We comment on the results, mirroring, and enriching, the observations regarding the cumulative nonlocal transport of section \ref{sec:4.4.1}.\\
{\bf Direct energy cascade solution $n_k^{\nu_{\omega}}$, Fig.~\ref{nonlocaltrans}(c)}
\begin{itemize}
    \item The density of energy transport is negative for $K<^\omega K_{\rm min}<^\omega K_{0}$, and positive for $K>^\omega K_{\rm min}$. Now, it becomes clear that the ``return flow'' (i.e. the inverse transfer in the direct energy cascade solution) is for wave numbers in which the cumulative nonlocal transport has negative tendency (and not negative value). Still, it highlights that wave numbers with large scale separation (larger than a factor $\/^\omega K_{\rm min}\simeq 4.5$) are subject to inverse energy transport.
    \item In the wave action density, the inverse-transport character is more pronounced. For action, all wave numbers smaller than $^\omega K_0$ undergo inverse transport, balancing exactly the nonlocal transport density of wave numbers greater than $^\omega K_0$ so that at $K=1$ the cumulative transport vanishes.
\end{itemize}
{\bf Inverse action cascade solution $n_k^{\nu_{1}}$, Fig.~\ref{nonlocaltrans}(d)}
\begin{itemize}
    \item In the inverse action cascade, we already know that the energy transport density must balance to zero when integrated. However, it is interesting to notice that now it is energy that undergoes a ``counter-flow'' direct transport for the more local transfers at wave numbers larger than about 0.55. Indeed, this information was already visible in the positive tendency of the cumulative transport in Fig.~\ref{nonlocaltrans}(b).
    \item On the other hand, for the same solution the action density transport is always negative, so that apparently a counter-flow is not present in this case for this conserved quantity.
\end{itemize}

\begin{figure}
    \hspace{-0.2cm}    \subfloat[]{\includegraphics[scale=\picscaletwo]{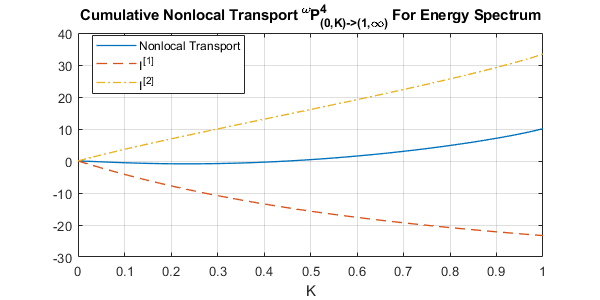} \label{NonlocalEnergyLeftRightB}}
    \hspace{-.5cm}
       \subfloat[]{\includegraphics[scale=\picscaletwo]{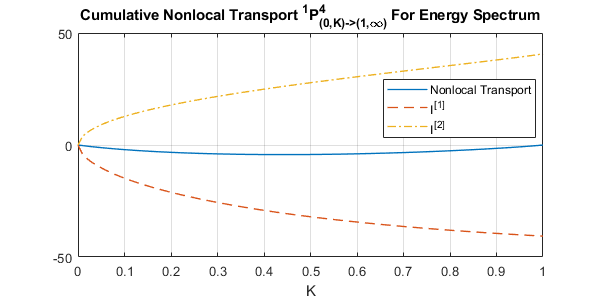}} \label{NonlocalActionLeftRightB}\\
       \hspace{-15cm}
    \subfloat[]
    {\includegraphics[scale=\picscaletwo]{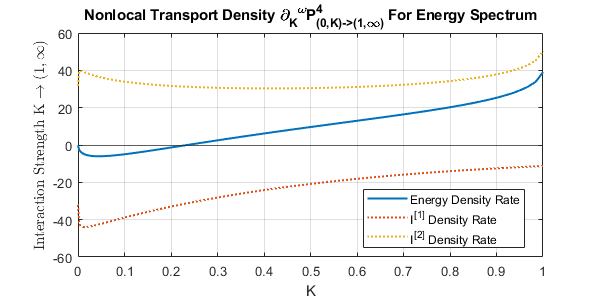}} \label{NonlocalEnergySplitDensity}
    \hspace{-.5cm}
    \subfloat[]{\includegraphics[scale=\picscaletwo]{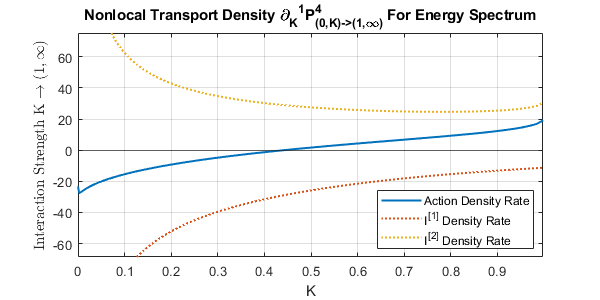}} 
    \label{NonlocalActionSplitDensity}   
    \caption{All figures refer to the {\bf energy cascade} Kolmogorov-Zakharov solution. The nonlocal transport cumulative distributions are found in (a) for energy and in (b) for action. The respective nonlocal transport density distributions are found in (c) for energy and in (d) for action. }\label{fig:5}
\end{figure}

\subsubsection{Splitting the transports: direct and inverse contributions}

The nonlocal transport analysis in the above sections is performed for the net transfers which come from balancing positive and negative contributions in the RHS of \eqref{angleavMMT}. As detailed in \eqref{thing1}, we are actually able to distinguish the negative ($I^{[1]}$) and positive ($I^{[2]}$) contributions to the transfers, and analyze them independently. This allows us to gain deeper insight into the superposition of direct and inverse transports that may be hidden in the previous analysis (because superposed and with opposite sign). For simplicity, here we only focus on the energy cascade Kolmogorov-Zakharov solution. The result of the splitting is shown in Fig.~\ref{fig:5}, where we show plots of nonlocal transport cumulative distribution for the positive (direct) and negative (inverse) contributions (Fig. \ref{fig:5}(a) for energy and Fig. \ref{fig:5}(b) for action), and their respective density distributions (Fig. \ref{fig:5}(c) for energy and Fig. \ref{fig:5}(d) for action). The contribution from net transport, given by summing the positive and negative contributions, is also shown (same as in Fig.~\ref{nonlocaltrans}(a)). We note some important facts:
\begin{itemize}
    \item For both energy and action transport, over half of its total (values at $K=1$ in (a)) is concentrated at wave numbers smaller than about $0.5$ (Fig.~\ref{fig:5}(a)), and smaller than about $0.2$ for action (Fig.~\ref{fig:5}(b)). This would suggest the median scale separation of the transfers to be about a factor of $2$ for energy, and about a factor of $5$ for action. We will inquire the metrics of locality more precisely in the next section.
    \item The magnitude of the direct and inverse transfers are on average one order of magnitude larger than the net transfer. This is a very important observation about the physical system that is hidden when only the net flux is computed. Within the energy and actions cascades, energy and action are being transferred vigorously in both directions. For energy, the direct transfers are qualitatively more local, while the inverse transfers are qualitatively more nonlocal (yellow curve in panel (c) more concentrated to the right and yellow curve to the left). Indeed, the net of the two contributions is the positive flux of the direct energy cascade.
    \item The simultaneous existence of positive and negative contributions should not be a surprise looking at how the four-wave interactions take place, as represented in Fig.~\ref{schematicresman}. Each individual interaction is characterized by a direct and an inverse simultaneous contributions.
    \item Due to the observations above, a rigorous study of locality must therefore be based on the properties of the direct and inverse transfers separately. If this is not done, cancellations in the net flux are likely to hide or misrepresent the actual locality properties of the system.
    \end{itemize}

\subsection{Local Transport}

\subsubsection{(Normalized) cumulative local transport}
{{We now introduce a local flux characteristic. Specifically we introduce }}
the normalized local transport $\mathcal{P}(a,b)$, which depends on one more free parameter. Let $a<k<b$ be wave numbers, then the normalized local transport across fixed wave number $k$ is given by

\begin{equation}
\mathcal{P}(a,b)=\frac{{^{\rho} P^4_{(a,k)\to (k,b)}}}{{^{\rho}P^4_k}};\quad a<k<b\label{normalizedpartialflux}
\end{equation}

The normalized local transport is a fundamentally two-dimensional quantity. The normalization permits clearer interpretation of the color map figures. $\mathcal{P}(a,b)$ visualizes the conserved quantity flow from an alternate perspective to nonlocal transport; rather than considering the flow through a gap by removing the most local interacting wave numbers from the receiving set, we remove the most \textit{nonlocal} interactions and study the flow across a point considering interactions between adjacent sets.

The results are shown in figure \ref{Fluxab}.  Note that the local transport is symmetric across the  $a b=1$ curve. This can
be verified  analytically by equations \eqref{prop1leftsplit} and \eqref{prop2rightsplit},  \eqref{ScaleInvar}:

\begin{figure}
\hspace{-.5cm}
    \subfloat[]{\includegraphics[scale=.6]{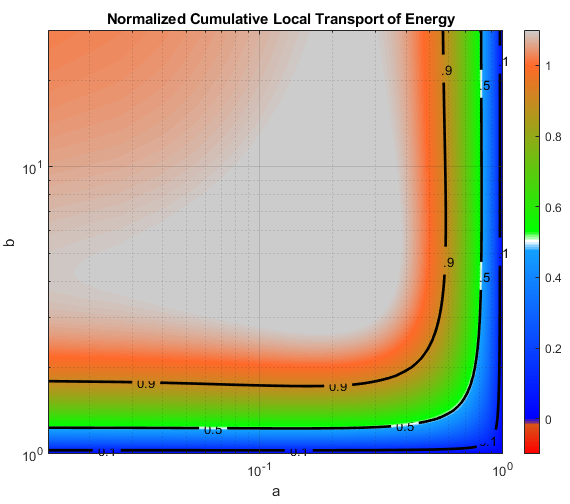} \label{normloctrans_en}}
    \hspace{-1cm}
    \subfloat[Ch]{\includegraphics[scale=.605]{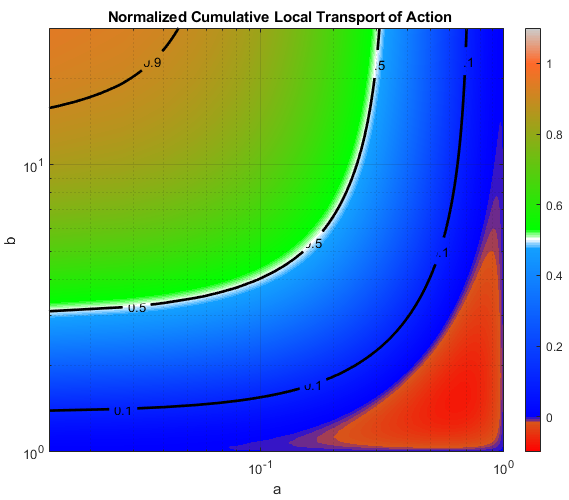} \label{normloctrans_ac}}   
    \caption{The plot of $^{\rho}P^4_{(a,1)\to (1,b)}/{^{\rho}P^4_1}$ for (a) energy transfer at the energy Kolmogorov-Zakharov solution $n_k^{\nu_{\omega}}$ and for (b) action at the action Kolmogorov Zakharov solution $n_k^{\nu_1}$. Plotted in log-log scale with $a<1<b$. The left image represents normalized partial energy flux and the right image represents normalized partial action flux. The level curves at $.9,\:.5,\:.1$ represent the curve in a-b space at which the transfer obtains that percentage of the total flux. Note that the energy flux obtains the total energy flux for finite $a$ and $b$, while the action does not up to the bounds of the figure.}\label{Fluxab}

\end{figure}

\begin{equation}\label{FlippingEquality}
\begin{aligned}
&^{\rho}P^4_{(a,1)\to(1,b)}\\
&={P^4_{(0,1)\to(1,\infty)}}-{P^4_{(0,1)\to(b,\infty)}}- \left({^{\rho}P^4_{(0,a)\to(1,\infty)}}-{^{\rho}P^4_{(0,a)\to(b,\infty)}}\right)\\
&={P^4_{(0,1)\to(1,\infty)}}-{P^4_{(0,1/b)\to(1,\infty)}}- \left({^{\rho}P^4_{(0,1)\to(1/a,\infty)}}-{^{\rho}P^4_{(0,1/b)\to(1/a,\infty)}}\right)\\
&={^\rho P^4_{(1/b,1)\to (1,1/a)}}
\end{aligned}
\end{equation}

In Fig.~\ref{Fluxab}(a), the white-ish region (with normalized local transport larger than 1, mapping into the negative nonlocal transport regions in the above sections) signals those combinations of wave numbers subject to net inverse energy transfer in the energy Kolmogorov-Zakharov solution, which are wave number pairs subject to large scale separation (i.e. $a$ small and $b$ large, upper right of the plot). This deserves further explanation below.

In Fig.~\ref{Fluxab}(b), related to the action Kolmogorov-Zakharov solution, the most local (i.e. both $a$ and $b$ close to 1, lower left of the plot) transfers of action are actually showing a net transfer countering the inverse action cascade, in the red region in the plot. Note that this counterflow was not visible in the one-dimensional nonlocal transport distributions analyzed above. This leads us to the important observation that the inverse action cascade is driven by nonlocal interactions, and has a direct highly local return mechanism - specular to what happens to energy in the energy cascade. These features will be further explored in the next section, in which we will take partial derivatives of the local transport allowing for better intuition.

\subsubsection{Further analysis of cumulative local transport of energy in the energy Kolmogorov-Zakharov spectrum}

\begin{figure}
\begin{center}
\includegraphics[scale=\picscale]{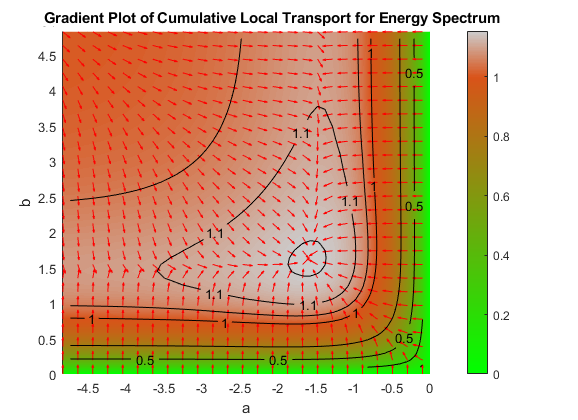}
    \caption{The gradient of the normalized local transport of energy. The gradient is normalized for visibility. Arrows pointing towards the fourth quadrant indicate both $a\to (1,b)$ and $(a,1)\to b$ are inverse (upper-left part of the figure).}\label{fluxenergygradient}
\end{center}
\end{figure}

The cumulative local transport of energy for the energy stationary solution is found in Fig. \ref{normloctrans_en}. 
We mentioned that  the region where the local transport is greater than the total flux $^{\omega}P^4_1$ indicates energy transferred upscale against the direct cascade. To more precisely understand what this area indicates, we plot the gradient (Fig. \ref{fluxenergygradient}) of the normalized local transport, which by the reasoning of \eqref{wnatoB} describes the individual interactions as

\begin{equation}
\nabla_{(a,b)} {^{\omega}P^4_{(a,1)\to (1,b)}}=\left[-{^{\omega}P^4_{a\to (1,b)},{^{\omega}P^4_{(a,1)\to b}}}\right]
\end{equation}

We see a fairly distinct set of relationships governing the forward and backward transfers of energy. When both $a$ and $b$ are close to the fixed wave number $k=1$, both the transfer $a\to (1,b)$ and $(a,1)\to b$ are positive, further confirming that the most local interactions drive the forward cascade. We can also note that the energy return mechanism is quite dominant, with forward point-set interactions restricted to $a\gtrapprox.1$ or $b\lessapprox 10$.

\subsubsection{Point-point transport densities}
For greater detail, we can further press the reasoning of \eqref{wnatoB}. Note that the transfer $a\to B$ retains the property \eqref{prop2rightsplit}, so that $P_{a\to B\cup C} = P_{a\to B}+P_{a\to C}-P_{a\to B\cap C}$. Hence if we choose $B=[b,b']$ where $1<b,b'$ we see

\begin{equation}
\begin{aligned}
\frac{P_{a\to B}}{b'-b}&=\frac{P_{a\to (b,\infty)}-P_{a\to (b',\infty)}}{b'-b}\\
\int_0^{\infty}\delta(k-a)\int_1^{\infty}\frac{\chi_B(k_1)}{b'-b}\int dk_{23}\rho_k{^{\rho}W_{12}\cali^{12}_{3k}}&=-\frac{P_{a\to (b',\infty)}-P_{a\to (b,\infty)}}{b'-b}\\
\xrightarrow{b'\to b} \int_0^{\infty}\delta(k-a)\int_1^{\infty}\delta(k_1-b)\int dk_{23}\rho_k{^{\rho}W_{12}\cali^{12}_{3k}}&=-\partial_b P_{(a,1)\to (1,b)}
\end{aligned}
\end{equation}

The above can be summarized as the point-point transport density from $a$ to $b$ as

\begin{equation}\label{pointotpointdef}
^{\rho}P^4_{a\to b} = -\partial_a\partial_b{^{\rho}P^4_{(a,1)\to (1,b)}}\,.
\end{equation}

\begin{figure}
    \hspace{-1cm}
       \subfloat[]{\includegraphics[scale=.4]{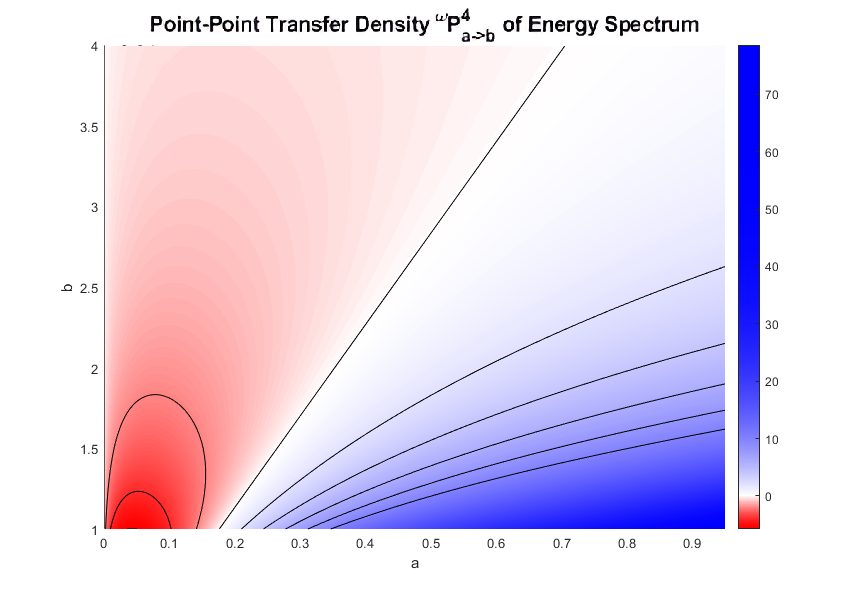}
    \label{individualtrans_en}}
    \hspace{-.1cm}
    \subfloat[]{\includegraphics[scale=.4]{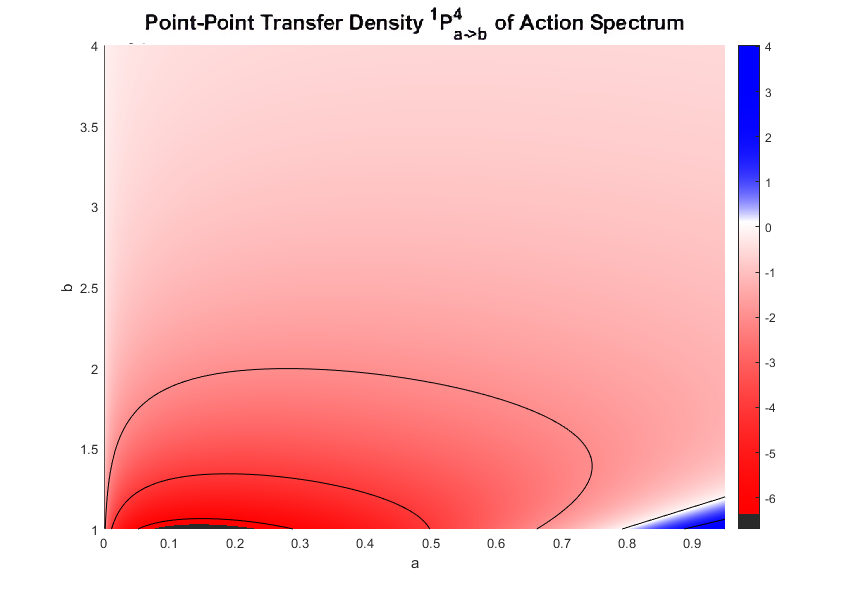}\label{individualtrans_ac}}

    \caption{The point-point transfer density from $a\to b$ where $a<1<b$. (a) We plot the energy transfer density for the energy cascade solution and (b) the action transfer density for the action cascade solution.}\label{IndividualTransferStuff}   

\end{figure}

We plot this weighted evaluation of the collision integrand in Fig. \ref{individualtrans_en} for energy in the energy cascade solution and in Fig. \ref{individualtrans_ac} for action in the action cascade solution.\\
{\bf Point-point energy transfers in the energy cascade solution (Fig. \ref{individualtrans_en})}:
Our observations are made the most concrete when observing the point-point transport densities: we may summarize all of our observations about the energy transfers underlying the energy cascade with this image.
\begin{itemize}
    \item The cascade of energy is forward, and this forward flow is driven by fairly local steps. The interactions of greatest magnitude are those which are linearly closest together in wave number space.
    \item There is a nonlocal energy return mechanism where inverse interactions dominate the forward ones. In the point-point transfer calculation, this manifests as a line across which the sign of net interaction between points changes. 
\end{itemize}
    {\bf Point-point action transfers in the action cascade solution (Fig. \ref{individualtrans_en})}:
    \begin{itemize}
        \item The cascade of action is inverse, and this inverse flow includes very nonlocal steps. Very nonlocal interaction have large magnitude.
    \item There is a highly local forward return mechanism where forward interactions dominate the inverse ones, which is quite large in magnitude. This is apparent in the lower-right corner of the picture.
    \end{itemize}

\section{Locality Quantifications}

\subsection{Locality Curves and Metrics of Locality}

What proportion of a sending set which contains the wavenumbers most local to the receiving set is required to capture p\% of the total flow?. To 
answer this question, we use  the local transport through fixed wave number $k$ given by $^{\rho}P^4_{(K,k)\to (k,\infty)}$. This quantity describes the transfer rate from a finite set to the dissipative (arbitrarily large wave numbers) region. The parameter $K$ controls the locality in wave number space of the sending set to the output region. To set up a system of measurement, we interpret the normalized local transport as the proportion of the total flow which originates from a set $(K,k)$ as the locality is increased ($K\to k$). We term this the locality curve, written\\
\begin{align}
&{^{\rho}\mathcal{L}}(K,k)=\frac{^{\rho}P^4_{(K,k)\to (k,\infty)}}{^{\rho}P^4_{(0,k)\to (k,\infty)}},
\:\text{ with }n_k^{\nu_\rho}\label{localitycurveformula}
\end{align}

The locality curve ${^{\rho}\mathcal{L}}(K,1)$ (Fig. \ref{LC}) is a function of $K$ which represents the proportion of $^{\rho}P^4_1$ captured by the transfer from a subset $^{\rho}P^4_{(K,1)\to (1,\infty)}$. Practically, it is a function which relates the proportion of $(0,1)$ covered by $(K,1)$ with the proportion of $^{\rho}P^4_{(K,1)\to (1,\infty)}$ with respect to the flux $^{\rho}P^4_1$. We will define $K_{p}$ as the solution to ${^{\rho}\mathcal{L}}(K,1)=p$, determining the set $(K_{p},1)$ at which the proportion $p$ of the total flux is realized.

\begin{figure}
\hspace{-.5cm}
    \subfloat[]{\includegraphics[scale=\picscale]{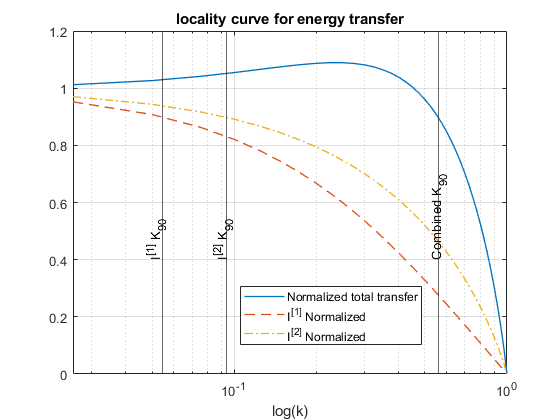}\label{LCEnL}}
    \hspace{-1cm}
    \subfloat[]{\includegraphics[scale=\picscale]{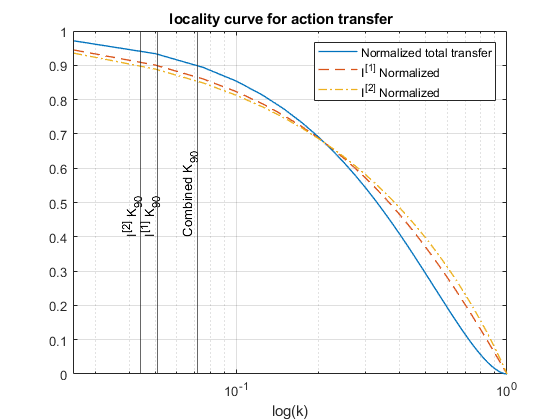} \label{LCAcL}}
    
    \caption{The locality curves for fixed forward cascade solution. In each case, $K_{.9}$ refers to the set $(K_{.9},1)$ (a) The locality curve with energy cascade solution. $I^1<0<I^2$ curves captured analogously. Note that $K_{1}={^{\omega}K_0}$ for the combined transfer, but the constituent curves for $I^{[i]}$ are not captured locally. (b) The locality curve for energy cascade solution.}\label{LC}    

\end{figure}

Under this paradigm, the curve skewing towards $K=0$ implies less local interactions drive the flux $^{\rho}P^4_{1}$. In  Fig. \ref{LC}, we plot the value $K_{.9}$, corresponding to $K$ at which $(K_{.9},1)$ captures 90\% of the total quantity transferred from $(0,1)$ into $(1,\infty)$. We would then state that smaller magnitudes of $K_{.9}$ correspond to greater nonlocality of the flux $^{\rho}P^4_{(0,1)\to(1,\infty)}$. Note that we additionally plot locality curves for the constituent inverse (negative) transfer $I^{[1]}$, and the constituent forward (right) transfer $I^{[2]}$. These are calculated similarly to the full curves, where rather than normalizing using the total flux, we normalize using $I^{[i]}_{(0,1)\to(1,\infty)}$.

We note that for the flux of energy at the constant energy flux solution (Fig. \ref{LCEnL}), unity is obtained at $K_1\approx .45$, in correspondence with its counterpart figure \ref{NonlocalEnergyLeftRight}. We note that this property is not enjoyed by the forward/backward decomposition $I^{[i]}$ as these constituent integrals are strictly single-signed. Thus the constituent interactions making up the energy have $K_{.9}$ of approximately one order of magnitude less than the same metric for the full locality curve.

The metric $K_{.9}$ for action transfer at the constant waveaction flux solution indicates far greater nonlocality under our paradigm than the energy locality, in accordance with our observations in section 4. The constituent transfer rates are comparable in locality to the total flux of action.

Note that if one wants to exclude direct connection with a forcing or dissipative region, and take the $10\%$ area in the tale as a cutoff of what can be neglected, we would need to ensure a scale separation of a factor of $1/K_{.9}$. In doing so, the constituent transfers (independent positive and negative contributions) are the meaningful ones to use, leading to a separation factor of more than an order of magnitude for both cascades (indeed, quite a large factor!). This fact might explain why in the numerical simulations of the MMT model one needs to reach extremely large box size in order to be able to observe a clean Kolmogorov-Zakharov solution that is independent of the characteristics of the forcing and dissipation~\parencite{buhlerMMT2023,hrabski2022properties}.

\subsection{Median Steps of Energy}

Let us examine a calculation which is related to the normalized local transport \eqref{normalizedpartialflux}. We introduce the median step scaling of energy, $s(a)$, of an interval $(aK,K)$ for some $a\in (0,1)$ as the implicit solution to 
\begin{equation}\label{StepSize}
\frac{^\rho P_{(aK,K)\to (K,K/s(a))}}{^\rho P_{(aK,K)\to (K,\infty)}}=.5\text { at }n_k^{\nu_{\rho}}
\end{equation}

That is, given a set $A=(aK,K)$ we choose $B(A)=(K,K/s(a))$ to be the set such that the $B$ captures 50\% of the energy leaving $A$ to the right. Note that by the scale invariance property \eqref{ScaleInvar}, when considering Kolmogorov-Zakharov solutions, it is sufficient to perform all calculations for $K=1$, making $s(a)$ scale invariant.

We put forth the median step size as a rudimentary method for examining the dynamics of a cascade state. If we were to "follow" the energy contained in $A$ iteratively, to $B_1=B(A)$, as a first step, $B_2=B(B(A))$ as a second step, and beyond, we can get an intuitive approximation for how the energy in $A$ propagates through an interval $\bigcup\limits_{i=1}^n B_i$. We could then treat the median steps as a measure of distance and time, insomuch as every two units of time the amount of energy transferred ${^\omega}P^4_{B_i\to B_{i+1}}$ will be equal to the total local transport ${^\omega}P^4_{B_i\to (k>b\in B_i,\infty)}$.

We will thus study $s(a)$ iteratively. Given $a$ and $K$, we define $s_1=b(a)$ as the solution to \eqref{StepSize}. Likewise, we define $s_2$ as the solution for the starting interval $[K,K/s_1]$. This equation can be rewritten by the scale invariance property \eqref{ScaleInvar} to become

\begin{equation}\label{StepSizeScaled}
\frac{^\rho P_{(K,K/s_1)\to (K/s_1,K/(s_1s_2))}}{^\rho P_{(K,K/s_1\to (K/s_1,\infty)}}=\frac{^\rho P_{(s_1K,K)\to (K,K/s_2)}}{^\rho P_{(s_1K,K)\to (K,\infty)}}=.5
\end{equation}

\begin{equation}
\begin{aligned}
\left(\frac{K}{s_1\cdot ... \cdot s_{i-1}},\frac{K}{s_1\cdot ... \cdot s_{i}}\right)\to 
\left(\frac{K}{s_1\cdot ... \cdot s_{i}},\frac{K}{s_1\cdot ... \cdot s_{i+1}}\right)\\
=\frac{1}{s_1\cdot ... \cdot s_{i}}\left[\left(s_iK,K\right)\to (K,K/s_{i+1})\right]
\end{aligned}
\end{equation}

\newcommand{\picscaletree}{.5}

\begin{figure}
\hspace{0cm}
    \subfloat[]{\includegraphics[scale=\picscaletree]{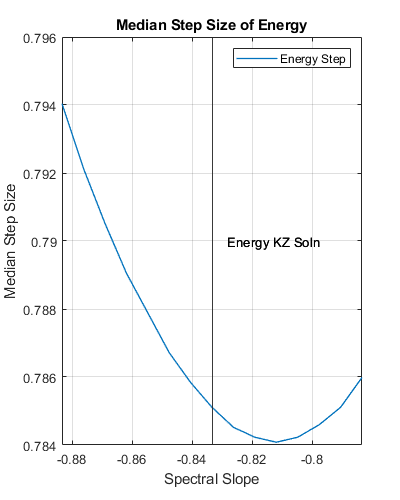}\label{MedStepEn}}
    \hspace{-.5cm}
    \subfloat[]{\includegraphics[scale=\picscaletree]{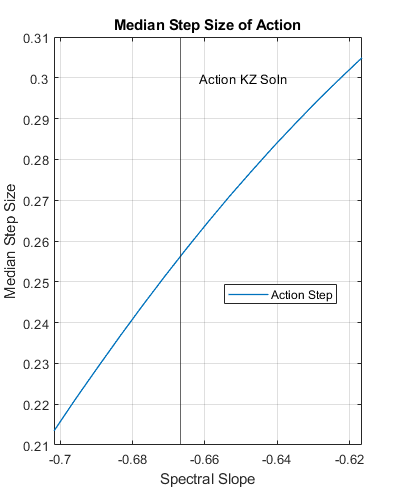} \label{MedStepAc}}
    
    \caption{The median step of energy and action as well as their sign decomposition $I^{[i]}$. Step size plotted as a function of spectral power law in an interval such that the direction of the denominator remains consistent.}\label{MedSteps}    

\end{figure}

Hence, $s_{i+1}=s(s_i)$ and by the scaling property we recover a dynamical system. $s(a)=a$ is an attracting fixed point so iteration will lead $s(a)\to a$. We then approximate the width of an interval $I=[i_1,i_2]$ in terms of median steps as $\text{steps}_{\rho}(I)=\log_a\left(\frac{i_1}{i_2}\right)$ where $a$, the "natural" median step is the solution to

\begin{equation}\label{naturalstep}
\frac{^{\omega}P^4_{(aK,K)\to \left(K,\frac{K}{a}\right)}}{^{\omega}P^4_{(aK,K)\to (K,\infty)}}=.5\text{ at }n_k^{\nu_{\omega}}
\end{equation}

We may further note that using properties \eqref{equalinandout}, \eqref{FlippingEquality}, \eqref{ScaleInvar}, and \eqref{prop3flip}, the median step definition may be written for $b>0$ in a form relevant to the inverse cascade of action (by capturing flow from right to left)

\begin{equation}\label{actionnaturalstep}
\frac{^{1}P^4_{\left(K,bK\right)\to (K/b,K)}}{^{1}P^4_{(K,bK)\to (0,K)}}=.5\text{ at }n_k^{\nu_{1}}
\end{equation}

Due to its definition, the factor that the median transport implies in terms of scales is given by $1/s$.
In Fig.~\ref{MedSteps}, we show the step size as a function of the spectral slope, in panel (a) for energy and in panel (b) for action. For energy the step size is smaller than for action, as the blue curve in the plots is closer to unity.

The characterization of  a step size allows to intuitively think 
about cascades as a dynamical system, where, on average, the
wave numbers jump by a step size: $k_{i+1}= s k_i$, where $k_i$ is 
a wave number at a step $i$, and $k_{i+1}$ is an average wave number at a step $i+1$ after the scattering event occured. We have determined
numerically (Fig. \ref{MedSteps}) that for the direct energy cascade the step size is $1.3$, so that $k_{i+1}\simeq 1.3 k_i$ and the wave numbers grow on average. 
For the inverse action cascade we found the step size to be about  
$1/4$, so that $k_{i+1}\simeq k_i/4$. Therefore the wave numbers 
decrease, on average. Moreover, we observe that the energy cascade
is much more local than the wave action cascade, which has much larger
jumps.

\section{Conclusions}\label{Conclusions}
In this manuscript we have  onstructed a general theory to investigate the fluxes of 
conserved quadratic invariants in the wave turbulence. We have considered three-wave,
four-wave, five-wave and six-wave interacting systems. We studied in details the 
celebrated Majda-McLaughin-Tabak system to demonstrate effectiveness and versatility of our
method. 

To summarize, we have 

\begin{itemize}
\item We have established a systematic formalism that unpacks the information in the collision integral of a wave kinetic equation in terms of sum of detailed balanced interactions. This allows us to restrict the transfers of any sign-definite conserved quantity in the system between any pair of arbitrary subsets of the spectral domain. We have started from illustrating our theoretical method for the four-wave case, which we have tackled in section 2. We than have repeated for cases of the three-, five-, and six-wave kinetic equations in section 3.

\item We have shown that in the isotropic case, the flux across a sphere that can be commonly written by interpreting the wave kinetic equation as a continuity equation is a particular case of our transfer formula, in the case of adjacent sets. 

\item Remarkably, the fluxes in the Kolmogorov-Zakharov solutions can be easily computed via our formula as integrals of well-defined nonzero quantities, without the need to regularize an indeterminate limit, and in a way that unveils how nonlocal the underlying transfers actually are.

\item We have then applied our formalism to the renowned Majda-McLaughlin-Tabak model, to analyze the properties of the transfers of energy and action at its Kolmogorov-Zakharov cascade solutions.

\item We have found that both the energy and the action cascades show ``return flow'' systems: subsets of the spectral transfers in which the flow of conserved quantities opposes the direction of their cascade. In each case, however, the return mechanism operates under a differing paradigm: the direct energy cascade is driven by highly local forward interactions, and returns energy backward by far more nonlocal inverse interactions. The inverse action cascade flips this script, being driven by highly nonlocal interactions, with a return mechanism of very local steps sending action back towards higher wave numbers.

\item We have found that the net transfers are the residuals of much larger transfers that are simultaneously transferring energy upscale and downscale with nontrivial different distributions. We are able to single out these constituent transfers and to study their locality properties.

\item In section 5, we have analyzed the locality properties of the wave kinetic equation associated with the Majda-McLaughlin-Tabak model. The action cascade is indeed more nonlocal than the energy cascade (roughly by a factor of 3 in its median step size). Moreover, we have estimated a safety factor for not being directly dynamically coupled with the forcing and dissipation regions. This factor is larger than an order of magnitude for both cascades. Such a large scale separation may explain why in the numerical simulations of the Majda-McLaughlin-Tabak model one needs to reach extremely large system size to be able to observe a fully developed Kolmogorov Zakharov spectrum that does not depend on the type of forcing/dissipation~\parencite{buhlerMMT2023,hrabski2022properties}.

\item In general, we have defined transport cumulative and density distributions, which have allowed us to measure the main properties of the transfers systematically. Moreover, we have derived portable formulas that can be applied to study the transport properties of any wave turbulence system equipped with a wave kinetic equation.

\item The formulas that are derived in this manuscript are much more than a tool to analyze already known solutions. For instance, they apply directly to non-scale-invariant and nonisotropic systems, cases in which traditional approaches are not directly applicable -- e.g., see \cite{dematteis2024interacting} for a recent application to oceanic internal waves, which are neither scale invariant nor isotropic. 

\end{itemize}

In summary we have shown that the wave turbulence kinetic equation has a wealth of information about
the structure of energy fluxes. It is our hope that the formalism presented here might be a 
useful tool to uncover the mysteries hidden within the collision integral in generic wave turbulence systems.

\section*{Acknowledgments}
We thank Prof. Ashwani Kapila for discussion. Y.V.L. and G.D. acknowledge funding from the Simons Foundation through the Simons Collaboration on Wave Turbulence.

\printbibliography

\section{Appendix-Sign Decomposition of Isotropic Four Wave Kinetic Equation at Power Law Solutions}

We will consider the isotropic formulation of the transfer rate formula  \eqref{IsotropicFlux} with $\omega_k=|\textbf{k}|^{\alpha}=k^{\alpha}$. We will examine the structure of conserved quantity transfer in terms of leftward and rightward facing quartet interactions; that is, we interpret an interaction as being leftward if the sign of the transfer integrand is negative, and rightward in the positive case. Additionally,
it would be convenient to link these two cases to a partitioning of the resonant manifold.\\

There are two tools we will use to accomplish this. First, as we have
already noted, if $\omega_k$ is single-signed, and further if $\omega_k<\omega_3$ and $\omega_2<\omega_1$ then all viable resonant quartets under these conditions must satisfy,

\begin{equation}
\omega_1+\omega_2=\omega_3+\omega_k\implies 
\begin{cases}
\omega_1,\:\omega_2\in (\omega_k,\omega_3) & \omega_1<\omega_3\\
\omega_3,\:\omega_k\in(\omega_2,\omega_1) & \omega_1>\omega_3
\end{cases}
\end{equation} 

The alternate cases when $\omega_3<\omega_k$ and $\omega_1<\omega_2$ take analogous forms by the symmetry of the resonance condition $\omega_1+\omega_2=\omega_3+\omega_k$.\\

Second, The sign of the collision integrand is determined solely from the interaction expression $1/n_k+1/n_3-1/n_1-1/n_2$, where $n_k=k^{\nu}=\omega_k^{-x}$; the relabel $\nu=-x/\alpha$ is for convenience. We will demonstrate that the structure of the resonant manifold above is sufficient to determine the sign of this interaction.\\ 

\noindent
\textbf{Proposition:}\label{Proposition} Let $n_k=|k|^{\nu}=\omega_k^{\nu/ \alpha}$. Then for $(k_1,k_2,k_3,k_4)$ satisfying $\omega_1+\omega_2=\omega_3+\omega_k$ and $\nu<-\alpha$, the interaction term $1/n_k+1/n_3-1/n_1-1/n_2$ is positive for $\omega_3,\omega_k\in(\omega_2,\omega_1)$ and negative for $\omega_1,\omega_2\in (\omega_k,\omega_3)$ up to symmetric relabeling, and so the transfer formula is, respectively, negative and positive on these regions.\\

\noindent
\textbf{Proof:} Again, for convenience we let $x=-\nu/\alpha$. We first consider the system of resonant condition and positive interaction expression, which corresponds to the transfer integrand being negative.

\begin{equation}\label{SignedResMan}
\begin{cases}
\omega_1+\omega_2-\omega_3-\omega_k=0\\
\omega_k^{x}+\omega_3^x-\omega_1^x-\omega_2^x>0
\end{cases}
\end{equation}

We consider the resonant quartets satisfying
$\omega_3,\:\omega_k\in (\omega_2,\omega_1)$. This implies an ordering
which lets us write the second equation as a difference of positive
values: $(\omega_k^x-\omega^x_2)-(\omega^x_1-\omega^x_3)>0$. By this
ordering, we can reduce the independent variables of the
system from five to three by rewriting the paired terms as scalings of
one another $\omega_k=a\omega_2$ and $\omega_1=b\omega_3$ with
$a,\:b>1$ as we consider nontrivial solutions. We obtain a three
parameter equation

\begin{equation}
\begin{cases}
\frac{\omega_2}{\omega_3}=\frac{b-1}{a-1}\\
\frac{\omega_2}{\omega_3}<\left(\frac{b^x-1}{a^x-1}\right)^{1/x}\implies \frac{b-1}{a-1}<\left(\frac{b^x-1}{a^x-1}\right)^{1/x} 
\end{cases}
\end{equation}

Note that since $\omega_2<\omega_3$, $b<a$. We outline an argument to determine which spectral power laws $x$ make this inequality true.\\ 

\noindent
\textbf{Lemma-} For $x>0$ and $1<b<a$, $f(x)=\frac{a^x-1}{b^x-1}$ is strictly decreasing.\\

\textbf{Proof:} We take the derivative,

\begin{equation}
(b^x-1)^2f'(x)=\frac{1-b^{-x}}{\ln(b)}-\frac{1-a^{-x}}{\ln(a)}
\end{equation}

Let $g(b)=\frac{1-b^{-x}}{\ln(b)}$. As $a>b$, the lemma $f'(x)<0$ holds so long as $g(b)$ is strictly decreasing. Taking the derivative with respect to $b$,
\begin{equation}
b^{x+1}\ln(b)g'(b)=\ln(b^x)-(b^x-1)<0
\end{equation}

\vspace{.5cm}
Thus $g(b)$ is strictly decreasing, so $(b^x-a)^2f'(x)=g(b)-g(a)<0$ and $f$ is a strictly decreasing function on the assumption $b<a$. $\square$ \\

The function $f(x)$ is thus smooth and strictly decreasing on $x>0$, and $\lim\limits_{x\to 0^+}f(x)=\frac{\ln(a)}{\ln(b)}$ is well defined. Finally, we need to show that the smooth function $u(x)=f(x)^{1/x}$ is strictly increasing for $x>0$. We again take a derivative, and because $f'(x)<0$,

\begin{equation}
u'(x)=\frac{f'(x)f(x)^{1-x}}{x^2}\left(\frac{x}{f(x)}-\ln(f(x))\right)<0\implies x>f(x)\ln(f(x))
\end{equation}

This statement is trivially true as $0<f(x)<1$ since $b<a$.

Thus, our inequality $\frac{b-1}{a-1}<u(x)$ has a constant on the LHS for a fixed quartet, and a strictly increasing function of $x>0$ on the RHS. It follows $\frac{b-1}{a-1}=u(x)$ is uniquely solved for $x=1$, so the inequality holds for $x>1$.\\

Thus, when $x>1$ and $\omega_3,\omega_k\in (\omega_2,\omega_1)\text{ or }(\omega_1,\omega_2)$, the interaction expression is positive and the transfer integrand is negative.\\

It follows directly by permuting the indices $3\leftrightarrow 2$ and $1\leftrightarrow k$ that the case of the negative interaction expression, or positive transfer 

\begin{align}
\omega_1+\omega_2-\omega_3-\omega_k=0\\
\omega_k^{x}+\omega_3^x-\omega_1^x-\omega_2^x<0
\end{align}

is satisfied when $\omega_1,\omega_2\in (\omega_3,\omega_k)\text{ or }(\omega_k,\omega_3)$. Hence we have partitioned the nontrivial solutions of the wave kinetic equation resonant manifold into regions of resonant quartets where interactions are positive and negative, or negative and positive in the case of the transfer formula when $n_k=k^\nu$ and $\nu<-\alpha$. $\blacksquare$\\

The reasoning in the above proof can be extended to other cases for the power law; However $\nu<-\alpha$ is sufficient for our investigation. We can note in figure \eqref{MatchFig} that for $\nu>-\alpha$, the two cascades share a sign, meaning they are not individually physically realizable and fall outside the perview of pure WT \parencite{NazBook}. For the MMT integrals in the previous sections, the Proposition determines the sign decomposition $I^{[1]}<0<I^{[2]}$.\\

We interpret the regions of negative transfer as constituting left-moving, or higher to lower wave number interactions and the regions of positive transfer as right-moving or, or from lower to higher wave numbers.

\end{document}